%
%
%
%
\def\unredoffs{} \def\redoffs{\voffset=-.40truein\hoffset=-.40truein}
\def\speclscape{}
%
%
%
%
\newbox\leftpage \newdimen\fullhsize \newdimen\hstitle \newdimen\hsbody
\tolerance=1000\hfuzz=2pt
\catcode`\@=11 
\def\bigans{b }
\def\answ{b }
\ifx\answ\bigans\message{(This will come out unreduced.}
\magnification=1200\unredoffs\baselineskip=16pt plus 2pt minus 1pt
\hsbody=\hsize \hstitle=\hsize 
\else\message{(This will be reduced.} \let\l@r=L
\magnification=1000\baselineskip=16pt plus 2pt minus 1pt \vsize=7truein
\redoffs \hstitle=8truein\hsbody=4.75truein\fullhsize=10truein\hsize=\hsbody
\output={\ifnum\pageno=0 
  \shipout\vbox{\speclscape{\hsize\fullhsize\makeheadline}
    \hbox to \fullhsize{\hfill\pagebody\hfill}}\advancepageno
  \else
  \almostshipout{\leftline{\vbox{\pagebody\makefootline}}}\advancepageno
  \fi}
\def\almostshipout#1{\if L\l@r \count1=1 \message{[\the\count0.\the\count1]}
      \global\setbox\leftpage=#1 \global\let\l@r=R
 \else \count1=2
  \shipout\vbox{\speclscape{\hsize\fullhsize\makeheadline}
      \hbox to\fullhsize{\box\leftpage\hfil#1}}  \global\let\l@r=L\fi}
\fi
%
\newcount\yearltd\yearltd=\year\advance\yearltd by -1900

\def\Title#1#2{\nopagenumbers\abstractfont\hsize=\hstitle\rightline{#1}%
\vskip 1in\centerline{\titlefont #2}\abstractfont\vskip .5in\pageno=0}
\def\Date#1{\vfill\leftline{#1}\tenpoint\supereject\global\hsize=\hsbody%
\footline={\hss\tenrm\folio\hss}}
%

\def\draftmode{\message{ DRAFTMODE }\def\draftdate{{\rm preliminary draft:
\number\month/\number\day/\number\yearltd\ \ \hourmin}}%
\headline={\hfil\draftdate}\writelabels\baselineskip=20pt plus 2pt minus 2pt
 {\count255=\time\divide\count255 by 60 \xdef\hourmin{\number\count255}
  \multiply\count255 by-60\advance\count255 by\time
  \xdef\hourmin{\hourmin:\ifnum\count255<10 0\fi\the\count255}}}
\def\nolabels{\def\wrlabeL##1{}\def\eqlabeL##1{}\def\reflabeL##1{}}
\def\writelabels{\def\wrlabeL##1{\leavevmode\vadjust{\rlap{\smash%
{\line{{\escapechar=` \hfill\rlap{\sevenrm\hskip.03in\string##1}}}}}}}%
\def\eqlabeL##1{{\escapechar-1\rlap{\sevenrm\hskip.05in\string##1}}}%
\def\reflabeL##1{\noexpand\llap{\noexpand\sevenrm\string\string\string##1}}}
\nolabels
%
\global\newcount\secno \global\secno=0
\global\newcount\meqno \global\meqno=1
\def\newsec#1{\global\advance\secno by1\message{(\the\secno. #1)}
\global\subsecno=0\eqnres@t\noindent{\bf\the\secno. #1}
\writetoca{{\secsym} {#1}}\par\nobreak\medskip\nobreak}
\def\eqnres@t{\xdef\secsym{\the\secno.}\global\meqno=1\bigbreak\bigskip}
\def\sequentialequations{\def\eqnres@t{\bigbreak}}\xdef\secsym{}
\global\newcount\subsecno \global\subsecno=0
\def\subsec#1{\global\advance\subsecno by1\message{(\secsym\the\subsecno. #1)}
\ifnum\lastpenalty>9000\else\bigbreak\fi
\noindent{\it\secsym\the\subsecno. #1}\writetoca{\string\quad
{\secsym\the\subsecno.} {#1}}\par\nobreak\medskip\nobreak}
\def\appendix#1#2{\global\meqno=1\global\subsecno=0\xdef\secsym{\hbox{#1.}}
\bigbreak\bigskip\noindent{\bf Appendix #1. #2}\message{(#1. #2)}
\writetoca{Appendix {#1.} {#2}}\par\nobreak\medskip\nobreak}
%
%
\def\eqnn#1{\xdef #1{(\secsym\the\meqno)}\writedef{#1\leftbracket#1}%
\global\advance\meqno by1\wrlabeL#1}
\def\eqna#1{\xdef #1##1{\hbox{$(\secsym\the\meqno##1)$}}
\writedef{#1\numbersign1\leftbracket#1{\numbersign1}}%
\global\advance\meqno by1\wrlabeL{#1$\{\}$}}
\def\eqn#1#2{\xdef #1{(\secsym\the\meqno)}\writedef{#1\leftbracket#1}%
\global\advance\meqno by1$$#2\eqno#1\eqlabeL#1$$}
%
\newskip\footskip\footskip14pt plus 1pt minus 1pt 
\def\footnotefont{\ninepoint}\def\f@t#1{\footnotefont #1\@foot}
\def\f@@t{\baselineskip\footskip\bgroup\footnotefont\aftergroup\@foot\let\next}
\setbox\strutbox=\hbox{\vrule height9.5pt depth4.5pt width0pt}
\global\newcount\ftno \global\ftno=0
\def\foot{\global\advance\ftno by1\footnote{$^{\the\ftno}$}}
%
\newwrite\ftfile
\def\footend{\def\foot{\global\advance\ftno by1\chardef\wfile=\ftfile
$^{\the\ftno}$\ifnum\ftno=1\immediate\openout\ftfile=foots.tmp\fi%
\immediate\write\ftfile{\noexpand\smallskip%
\noexpand\item{f\the\ftno:\ }\pctsign}\findarg}%
\def\footatend{\vfill\eject\immediate\closeout\ftfile{\parindent=20pt
\centerline{\bf Footnotes}\nobreak\bigskip\input foots.tmp }}}
\def\footatend{}
%
%
\global\newcount\refno \global\refno=1
\newwrite\rfile
\def\ref{[\the\refno]\nref}
\def\nref#1{\xdef#1{[\the\refno]}\writedef{#1\leftbracket#1}%
\ifnum\refno=1\immediate\openout\rfile=refs.tmp\fi
\global\advance\refno by1\chardef\wfile=\rfile\immediate
\write\rfile{\noexpand\item{#1\ }\reflabeL{#1\hskip.31in}\pctsign}\findarg}
\def\findarg#1#{\begingroup\obeylines\newlinechar=`\^^M\pass@rg}
{\obeylines\gdef\pass@rg#1{\writ@line\relax #1^^M\hbox{}^^M}%
\gdef\writ@line#1^^M{\expandafter\toks0\expandafter{\striprel@x #1}%
\edef\next{\the\toks0}\ifx\next\em@rk\let\next=\endgroup\else\ifx\next\empty%
\else\immediate\write\wfile{\the\toks0}\fi\let\next=\writ@line\fi\next\relax}}
\def\striprel@x#1{} \def\em@rk{\hbox{}}
\def\lref{\begingroup\obeylines\lr@f}
\def\lr@f#1#2{\gdef#1{\ref#1{#2}}\endgroup\unskip}
\def\semi{;\hfil\break}
\def\addref#1{\immediate\write\rfile{\noexpand\item{}#1}} 
\def\startrefs#1{\immediate\openout\rfile=refs.tmp\refno=#1}
\def\xref{\expandafter\xr@f}\def\xr@f[#1]{#1}
\def\refs#1{\count255=1[\r@fs #1{\hbox{}}]}
\def\r@fs#1{\ifx\und@fined#1\message{reflabel \string#1 is undefined.}%
\nref#1{need to supply reference \string#1.}\fi%
\vphantom{\hphantom{#1}}\edef\next{#1}\ifx\next\em@rk\def\next{}%
\else\ifx\next#1\ifodd\count255\relax\xref#1\count255=0\fi%
\else#1\count255=1\fi\let\next=\r@fs\fi\next}
%

%
\newwrite\ffile\global\newcount\figno \global\figno=1
\def\fig{fig.~\the\figno\nfig}
\def\nfig#1{\xdef#1{fig.~\the\figno}%
\writedef{#1\leftbracket fig.\noexpand~\the\figno}%
\ifnum\figno=1\immediate\openout\ffile=figs.tmp\fi\chardef\wfile=\ffile%
\immediate\write\ffile{\noexpand\medskip\noexpand\item{Fig.\ \the\figno. }
\reflabeL{#1\hskip.55in}\pctsign}\global\advance\figno by1\findarg}
\def\xfig{\expandafter\xf@g}\def\xf@g fig.\penalty\@M\ {}
\def\figs#1{figs.~\f@gs #1{\hbox{}}}
\def\f@gs#1{\edef\next{#1}\ifx\next\em@rk\def\next{}\else
\ifx\next#1\xfig #1\else#1\fi\let\next=\f@gs\fi\next}
\newwrite\lfile
{\escapechar-1\xdef\pctsign{\string\%}\xdef\leftbracket{\string\{}
\xdef\rightbracket{\string\}}\xdef\numbersign{\string\#}}

\def\writestop{\def\writestoppt{\immediate\write\lfile{\string\pageno%
\the\pageno\string\startrefs\leftbracket\the\refno\rightbracket%
\string\def\string\secsym\leftbracket\secsym\rightbracket%
\string\secno\the\secno\string\meqno\the\meqno}\immediate\closeout\lfile}}
\def\writestoppt{}\def\writedef#1{}
\def\seclab#1{\xdef #1{\the\secno}\writedef{#1\leftbracket#1}\wrlabeL{#1=#1}}
\def\subseclab#1{\xdef #1{\secsym\the\subsecno}%
\writedef{#1\leftbracket#1}\wrlabeL{#1=#1}}
\newwrite\tfile \def\writetoca#1{}
\def\leaderfill{\leaders\hbox to 1em{\hss.\hss}\hfill}
\def\writetoc{\immediate\openout\tfile=toc.tmp
   \def\writetoca##1{{\edef\next{\write\tfile{\noindent ##1
   \string\leaderfill {\noexpand\number\pageno} \par}}\next}}}

%
\catcode`\@=12 
%
\edef\tfontsize{\ifx\answ\bigans scaled\magstep3\else scaled\magstep4\fi}
\font\titlerm=cmr10 \tfontsize \font\titlerms=cmr7 \tfontsize
\font\titlermss=cmr5 \tfontsize \font\titlei=cmmi10 \tfontsize
\font\titleis=cmmi7 \tfontsize \font\titleiss=cmmi5 \tfontsize
\font\titlesy=cmsy10 \tfontsize \font\titlesys=cmsy7 \tfontsize
\font\titlesyss=cmsy5 \tfontsize \font\titleit=cmti10 \tfontsize
\skewchar\titlei='177 \skewchar\titleis='177 \skewchar\titleiss='177
\skewchar\titlesy='60 \skewchar\titlesys='60 \skewchar\titlesyss='60
\def\titlefont{\def\rm{\fam0\titlerm}
\textfont0=\titlerm \scriptfont0=\titlerms \scriptscriptfont0=\titlermss
\textfont1=\titlei \scriptfont1=\titleis \scriptscriptfont1=\titleiss
\textfont2=\titlesy \scriptfont2=\titlesys \scriptscriptfont2=\titlesyss
\textfont\itfam=\titleit \def\it{\fam\itfam\titleit}\rm}
 \ifx\answ\bigans\else scaled\magstep1\fi
\ifx\answ\bigans\def\abstractfont{\tenpoint}\else
\font\abssl=cmsl10 scaled \magstep1
\font\absrm=cmr10 scaled\magstep1 \font\absrms=cmr7 scaled\magstep1
\font\absrmss=cmr5 scaled\magstep1 \font\absi=cmmi10 scaled\magstep1
\font\absis=cmmi7 scaled\magstep1 \font\absiss=cmmi5 scaled\magstep1
\font\abssy=cmsy10 scaled\magstep1 \font\abssys=cmsy7 scaled\magstep1
\font\abssyss=cmsy5 scaled\magstep1 \font\absbf=cmbx10 scaled\magstep1
\skewchar\absi='177 \skewchar\absis='177 \skewchar\absiss='177
\skewchar\abssy='60 \skewchar\abssys='60 \skewchar\abssyss='60
\def\abstractfont{\def\rm{\fam0\absrm}
\textfont0=\absrm \scriptfont0=\absrms \scriptscriptfont0=\absrmss
\textfont1=\absi \scriptfont1=\absis \scriptscriptfont1=\absiss
\textfont2=\abssy \scriptfont2=\abssys \scriptscriptfont2=\abssyss
\textfont\itfam=\bigit \def\it{\fam\itfam\bigit}\def\footnotefont{\tenpoint}%
\textfont\slfam=\abssl \def\sl{\fam\slfam\abssl}%
\textfont\bffam=\absbf \def\bf{\fam\bffam\absbf}\rm}\fi
\def\tenpoint{\def\rm{\fam0\tenrm}
\textfont0=\tenrm \scriptfont0=\sevenrm \scriptscriptfont0=\fiverm
\textfont1=\teni  \scriptfont1=\seveni  \scriptscriptfont1=\fivei
\textfont2=\tensy \scriptfont2=\sevensy \scriptscriptfont2=\fivesy
\textfont\itfam=\tenit \def\it{\fam\itfam\tenit}\def\footnotefont{\ninepoint}%
\textfont\bffam=\tenbf \def\bf{\fam\bffam\tenbf}\def\sl{\fam\slfam\tensl}\rm}
\font\ninerm=cmr9 \font\sixrm=cmr6 \font\ninei=cmmi9 \font\sixi=cmmi6
\font\ninesy=cmsy9 \font\sixsy=cmsy6 \font\ninebf=cmbx9
\font\nineit=cmti9 \font\ninesl=cmsl9 \skewchar\ninei='177
\skewchar\sixi='177 \skewchar\ninesy='60 \skewchar\sixsy='60
\def\ninepoint{\def\rm{\fam0\ninerm}
\textfont0=\ninerm \scriptfont0=\sixrm \scriptscriptfont0=\fiverm
\textfont1=\ninei \scriptfont1=\sixi \scriptscriptfont1=\fivei
\textfont2=\ninesy \scriptfont2=\sixsy \scriptscriptfont2=\fivesy
\textfont\itfam=\ninei \def\it{\fam\itfam\nineit}\def\sl{\fam\slfam\ninesl}%
\textfont\bffam=\ninebf \def\bf{\fam\bffam\ninebf}\rm}
%
%

\hyphenation{anom-aly anom-alies coun-ter-term coun-ter-terms}
\def\inv{^{\raise.15ex\hbox{${\scriptscriptstyle -}$}\kern-.05em 1}}

\def\Dsl{\,\raise.15ex\hbox{/}\mkern-13.5mu D} 
\def\dsl{\raise.15ex\hbox{/}\kern-.57em\partial}

 \def\Tr{{\rm Tr}}
\font\bigit=cmti10 scaled \magstep1
\def\lspace{\ifx\answ\bigans{}\else\qquad\fi}
\def\lbspace{\ifx\answ\bigans{}\else\hskip-.2in\fi} 
\def\boxeqn#1{\vcenter{\vbox{\hrule\hbox{\vrule\kern3pt\vbox{\kern3pt
    \hbox{${\displaystyle #1}$}\kern3pt}\kern3pt\vrule}\hrule}}}
\def\mbox#1#2{\vcenter{\hrule \hbox{\vrule height#2in
        \kern#1in \vrule} \hrule}}  
%
 \def\CO{{\cal O}} 
\def\CA{{\cal A}}   
\def\CL{{\cal L}} \def\CH{{\cal H}} \def\CI{{\cal I}} 
  \def\CD{{\cal D}} 
\def\e#1{{\rm e}^{^{\textstyle#1}}}

\def\darr#1{\raise1.5ex\hbox{$\leftrightarrow$}\mkern-16.5mu #1}
\def\ha{{1\over 2}}
\def\half{{\textstyle{1\over2}}} 
\def\roughly#1{\raise.3ex\hbox{$#1$\kern-.75em\lower1ex\hbox{$\sim$}}}

\def\np#1#2#3{Nucl. Phys. {\bf B#1} (#2) #3}
\def\pl#1#2#3{Phys. Lett. {\bf #1B} (#2) #3}

\def\pr#1#2#3{Phys. Rev. {\bf #1} (#2) #3}
\def\ap#1#2#3{Ann. Phys. {\bf #1} (#2) #3}

\def\cmp#1#2#3{Comm. Math. Phys. {\bf #1} (#2) #3}
\def\mpl#1#2#3{Mod. Phys. Lett. {\bf #1} (#2) #3}

\def\jhep#1#2#3{JHEP {\bf#1}(#2) #3}

\def\ijmp#1#2#3{Int.~J.~Mod.~Phys. {\bf #1} (#2) #3}
\def\atmp#1#2#3{Adv.~Theor.~Math.~Phys.{\bf #1} (#2) #3}
\def\ap#1#2#3{Ann.~Phys. {\bf #1} (#2) #3}
\def\IB{\relax\hbox{$\inbar\kern-.3em{\rm B}$}}
\def\IC{\relax\hbox{$\inbar\kern-.3em{\rm C}$}}
\def\ID{\relax\hbox{$\inbar\kern-.3em{\rm D}$}}
\def\IE{\relax\hbox{$\inbar\kern-.3em{\rm E}$}}
\def\IF{\relax\hbox{$\inbar\kern-.3em{\rm F}$}}
\def\IG{\relax\hbox{$\inbar\kern-.3em{\rm G}$}}
\def\IGa{\relax\hbox{${\rm I}\kern-.18em\Gamma$}}
\def\IH{\relax{\rm I\kern-.18em H}}
\def\IK{\relax{\rm I\kern-.18em K}}
\def\IL{\relax{\rm I\kern-.18em L}}
\def\IP{\relax{\rm I\kern-.18em P}}
\def\IR{\relax{\rm I\kern-.18em R}}
\def\IZ{\relax\ifmmode\mathchoice{
\hbox{\cmss Z\kern-.4em Z}}{\hbox{\cmss Z\kern-.4em Z}}
{\lower.9pt\hbox{\cmsss Z\kern-.4em Z}}
{\lower1.2pt\hbox{\cmsss Z\kern-.4em Z}}
\else{\cmss Z\kern-.4em Z}\fi}
\def\II{\relax{\rm I\kern-.18em I}}

\def\ndt{{\noindent}}

\def\sssec#1{\ndt$\underline{#1}$}

\def\CA{{\cal A}}

\def\CD{{\cal D}}
\def\CE{{\cal E}}

\def\CH{{\cal H}}
\def\CI{{\cal I}}

\def\CL{{\cal L}}

\def\CN{{\cal N}}
\def\CO{{\cal O}}

\def\CS{{\cal S}}

\def\CV{{\cal V}}

\def\p{\partial}
\def\pb{\bar{\partial}}


\def\mb{\bar{m}}
\def\nb{\bar{n}}

\def\zb{\bar{z}}

\def\Tr{{\rm Tr}}

\def\hn{{\hat n}}

\def\inbar{\,\vrule height1.5ex width.4pt depth0pt}

\font\cmss=cmss10 \font\cmsss=cmss10 at 7pt

\def\a{{\alpha}}
\def\ap{{\a}^{\prime}}

\def\d{{\delta}}

\def\e{{\epsilon}}
\def\z{{\zeta}}
\def\ve{{\varepsilon}}

\def\m{{\mu}}
\def\n{{\nu}}

\def\l{{\lambda}}

\def\t{{\theta}}
\def\o{{\omega}}
\def\nc{noncommutative\ }
\def\npt{non-perturbative\ }
\def\hp{\hat\partial}
\def\kk{{\kappa}}
\def\lref{\begingroup\obeylines\lr@f}
\def\lr@f#1#2{\gdef#1{\ref#1{#2}}\endgroup\unskip}
\lref\mtoda{P.~Etingof, I.~Gelfand, V.~Retakh,
``Factorization of differential operators, quasideterminants, and
nonabelian Toda field equations''
q-alg/9701008}
\lref\curtjuan{C.~G.~Callan, Jr., J.~M.~Maldacena, \np{513}{1998}{198-212},
hep-th/9708147}
\lref\bak{D.~Bak, \pl{471}{1999}{149-154}, hep-th/9910135}
\lref\baklee{D.~Bak, K.~Lee, hep-th/0007107}
\lref\moriyama{S.~Moriyama,
hep-th/0003231}
\lref\wadia{A.~Dhar, G.~Mandal and S.~R.~Wadia,
\mpl{A7}{1992}{3129-3146}\semi
A.~Dhar, G.~Mandal and S.~R.~Wadia, \ijmp{A8}{1993}{3811-3828}\semi
A.~Dhar, G.~Mandal and S.~R.~Wadia,  \mpl{A8}{1993}{3557-3568}\semi
A.~Dhar, G.~Mandal and S.~R.~Wadia,  \pl{329}{1994}{15-26}}

\lref\mateos{D.~Mateos, ``Noncommutative vs. commutative descriptions of
D-brane BIons'', hep-th/0002020}
\lref\mrs{S.~Minwala, M.~ van Raamsdonk, N.~Seiberg,
``Noncommutative Perturbative Dynamics'', hep-th/9912072}
\lref\nahm{W.~Nahm, \pl{90}{1980}{413}\semi
W.~Nahm, ``The Construction of All Self-Dual Multimonopoles
by the ADHM Method'', in ``Monopoles in quantum field theory'', Craigie et
al., Eds., World Scientific, Singapore (1982) \semi
N.J.~Hitchin, \cmp{89}{1983}{145}}
\lref\rs{M.~van Raamsdonk, N.~Seiberg, ``Comments of Noncommutative
Perturbative Dynamics'', hep-th/0002186, \jhep{0003}{2000}{035}}
\lref\k{M.~Kontsevich, ``Deformation quantization of Poisson
manifolds'', q-alg/9709040}
\lref\gms{R.~Gopakumar, S.~Minwala, A.~Strominger,
hep-th/0003160, \jhep{0005}{2000}{020}}
\lref\sst{N.~Seiberg, L.~Susskind, N.~Toumbas, hep-th/0005040}
\lref\sdual{R.~Gopakumar, S.~Minwala, J.~Maldacena, A.~Strominger,
hep-th/0005048\semi
O.~Ganor, G.~Rajesh, S.~Sethi, hep-th/00050046}
\lref\filk{T.~Filk, ``Divergencies in a Field Theory on  Quantum Space'',
\pl{376}{1996}{53}}
\lref\cf{A.~Cattaneo, G.~Felder, ``A Path Integral Approach to the Konstevich
Quantization Formula'', math.QA/9902090}

\lref\cds{A.~Connes, M.~Douglas, A.~Schwarz, \jhep{9802}{1998}{003}}
\lref\douglashull{M.~Douglas, C.~Hull, ``D-Branes and the noncommutative torus'', \jhep{9802}{1998}{008}, hep-th/9711165}
\lref\wtnc{E.~Witten, \np{268}{1986}{253}}
\lref\volker{V.~Schomerus, ``D-Branes and Deformation Quantization'',
\jhep{9906}{1999}{030}}
\lref\cg{E.~Corrigan, P.~Goddard, ``Construction of instanton and
monopole solutions and reciprocity'', \ap{154}{1984}{253}}

\lref\donaldson{S.K.~Donaldson, ``Instantons and Geometric
Invariant Theory", \cmp{93}{1984}{453-460}}

\lref\nakajima{H.~Nakajima, ``Lectures on Hilbert Schemes of
Points on Surfaces''\semi AMS University Lecture Series, 1999,
ISBN 0-8218-1956-9. }

\lref\neksch{N.~Nekrasov, A.~S.~Schwarz, hep-th/9802068,
\cmp{198}{1998}{689}}

\lref\freck{A.~Losev, N.~Nekrasov, S.~Shatashvili, ``The Freckled
Instantons'', {\tt hep-th/9908204}, Y.~Golfand Memorial Volume,
M.~Shifman Eds., World Scientific, Singapore, in press}

\lref\rkh{N.J.~Hitchin, A.~Karlhede, U.~Lindstrom, and M.~Rocek,
\cmp{108}{1987}{535}}

\lref\alexios{A.~Polychronakos, ``Flux tube solutions in noncommutative gauge theories'',
hep-th/0007043}
\lref\branek{H.~Braden, N.~Nekrasov, hep-th/9912019\semi
K.~Furuuchi, hep-th/9912047}

\lref\wilson{G.~ Wilson, ``Collisions of Calogero-Moser particles
and adelic Grassmannian", Invent. Math. 133 (1998) 1-41.}

\lref\gkp{S.~Gukov, I.~Klebanov, A.~Polyakov,
hep-th/9711112, \pl{423}{1998}{64-70}}
\lref\abs{O.~Aharony, M.~Berkooz, N.~Seiberg,
hep-th/9712117, \atmp{2}{1998}{119-153}}

\lref\abkss{O.~Aharony, M.~Berkooz, S.~Kachru, N.~Seiberg,
E.~Silverstein, hep-th/9707079, \atmp{1}{1998}{148-157}}

\lref\witsei{N.~Seiberg, E.~Witten, hep-th/9908142, \jhep{9909}{1999}{032}}
\lref\kinks{E.~Teo, C.~Ting, ``Monopoles, vortices and kinks in the
framework of noncommutative geometry'',
\pr{D56}{1997}{2291-2302}, hep-th/9706101}

\lref\manuel{D.-E.~Diaconescu, \np{503}{1997}{220-238}, hep-th/9608163}

\lref\genmnp{L.~Jiang,
``Dirac Monopole in Non-Commutative Space'', hep-th/0001073}
\lref\hashimoto{K.~Hashimoto, H.~Hata, S.~Moriyama,
hep-th/9910196, \jhep{9912}{1999}{021}\semi
A.~Hashimoto,
K.~Hashimoto, hep-th/9909202, \jhep{9911}{1999}{005}\semi
K.~Hashimoto, T.~Hirayama, hep-th/0002090}
\lref\hklm{J.~Harvey, P.~Kraus, F.~Larsen, E.~Martinec,
hep-th/0005031}

\lref\snyder{H.~S.~Snyder, ``Quantized Space-Time'', \pr{71}{1947}{38};
``The Electromagnetic Field in Quantized Space-Time'', \pr{72}{1947}{68}}
\lref\connes{A.~Connes, ``Noncommutative geometry'', Academic Press (1994)}

\lref\barsminic{I.~Bars, D.~Minic,
``Non-Commutative Geometry on a Discrete Periodic Lattice and Gauge Theory'',
hep-th/9910091}
\lref\grossnek{D.~Gross, N.~Nekrasov,
``Monopoles and Strings in Noncommutative Gauge Theory'', \jhep{0007}{2000}{034}, hep-th/0005204}
\lref\grossneki{D.~Gross, N.~Nekrasov, to appear}

\Title{\vbox{\baselineskip 10pt \hbox{PUPT-1945}
\hbox{ITEP-TH-39/00} \hbox{NSF-ITP-00-71} \hbox{hep-th/0007204}
{\hbox{ }}}} {\vbox{\vskip -30 true pt \centerline{DYNAMICS OF
STRINGS IN}
\smallskip
\smallskip
  \centerline{NONCOMMUTATIVE GAUGE THEORY}
\medskip
\vskip4pt }} \vskip -20 true pt \centerline{ David J.~Gross
$^{1}$, Nikita A.~Nekrasov $^{2}$}
\smallskip\smallskip
\centerline{ $^{1}$ \it Institute for Theoretical Physics,
University of California Santa Barbara CA 93106}
\centerline{$^{2}$ \it Institute for Theoretical and Experimental
Physics, 117259 Moscow, Russia} \centerline{$^{2}$ \it Joseph
Henry Laboratories, Princeton University, Princeton, New Jersey
08544} \medskip \centerline{\rm e-mail: gross@itp.ucsb.edu,
nikita@feynman.princeton.edu}
\bigskip

\centerline{\bf Abstract} \medskip
\noindent We continue our study
of solitons in \nc gauge theories and present an extremely simple
BPS solution of ${\CN} =4$ $U(1)$ \nc gauge theory in 4
dimensions, which describes    $N$ infinite D1
strings that pierce a D3 brane at various points, in the presence of a
background $B$-field in the Seiberg-Witten ${\ap} \to 0$ limit. We
call this solution the $N$-{\it fluxon}. For $N=1$ we calculate the
complete spectrum of small fluctuations about the fluxon and find
three kinds of modes: the fluctuations of the superstring in 10
dimensions arising from fundamental strings
attached to the D1 strings, the ordinary particles of the gauge theory in 4
dimensions and a set of states with  discrete spectrum, localized at the
intersection point
--- corresponding to fundamental strings stretched between the  D1
string and the D3 brane. We discuss the fluctuations about the
$N$-fluxon as well and derive explicit expressions for the
amplitudes of interactions between these various modes. We show
that translations in \nc gauge theories are equivalent to gauge
transformations (plus a constant shift of the gauge  field) and
discuss the implications for the translational zero modes of our
solitons. We also find the dyonic versions of $N$-fluxon, as well
as of our previous string-monopole solution. \Date{07/00}

\newsec{Introduction}
Field theories on noncommutative spaces \snyder\connes\
emerge as limits
of M theory compactifications
\cds\ or of string theory  with   D-branes in the presence of
a background Neveu-Schwarz $B$-field \douglashull\volker\witsei.
The interest in such theories is motivated by many analogies between
noncommutative gauge theories and
large $N$ ordinary non-abelian gauge theories \filk\mrs, and
by  the many features that noncommutative field
theories share  with open string  theory \wtnc\mrs\rs.

In this paper we continue the study \grossnek\ of \npt dynamical
objects in   noncommutative gauge theory, specifically four
dimensional gauge theories with an adjoint scalar field. Our paper
is organized as follows.

 In Section 2 we briefly review the setup
of \nc gauge theory. In \grossnek\ we discussed some general
features of these theories  and generalized Nahm's equations for
BPS solutions of the classical field equations to the
noncommutative theory. We solved these equations for the analogue
of a single monopole for noncommutative $U(1)$ theory. The
solution we constructed was nonsingular and sourceless, and
described a smeared monopole connected to a string-like flux tube.
We interpreted this string-monopole as the reflection of a D1
string attached to the D3 brane in the presence of a background
Neveu-Schwarz $B$-field. We calculated the tension of the string
and found precise agreement with that expected from the D1 string.
In Section 3 we briefly review this solution and then by deforming
it  construct an extremely simple classical BPS solution of \nc
$U(1)$ gauge theory with adjoint scalar field that describes an
infinite D1 string piercing the D3 brane, which we shall call the
{\it fluxon}. Then we find its generalization which describes $N$
D1 strings which pierce the D3 brane at various
points. This solution will be called the $N$-{\it fluxon}.

Despite being infinite these string-like solitons are not
translationally invariant---they depend on the specific point of
intersection---although the equations of motion are
translationally invariant. Thus the solitons we find are not
translationally invariant, although the theory is. Thus the
spectrum of small fluctuations should contain translational zero
modes. However, we find that in the \nc directions these modes are
essentially gauge transformations. Indeed, we show that in \nc
gauge theory translations are equivalent to (large) gauge
transformations plus shifts of the gauge field by a constant
amount.

The fluxon  solution is so simple that we are able to evaluate
explicitly the complete spectrum of fluctuations about the
soliton. This analysis is presented in Section 4. We find that the
fluctuating modes are those of fundamental strings. They fall into three
classes. These  correspond to light modes of fundamental
strings attached to the   D1 string and
to ordinary gauge, scalar and fermion particles that can thought
of as the light modes of fundamental strings attached to the D3
brane. In addition we find a set of modes with {\it
discrete} spectrum of energies that correspond to fundamental
strings that run between the D1 string and the D3 brane and are
localized near the point of intersection.

 In Section 5 we study the dynamics of the modes of the
fluxon - their propagation and their interaction with
the perturbative gauge particles and with the localized string
states.  In Section 6 we briefly generalize the discussion of fluctuations and
interactions to the case of the $N$-fluxon.

Finally, in Section 7, we   show that, having constructed
monopole-strings, we can also easily construct dyonic-strings.
These have a natural interpretation as the reflection in the gauge
theory of $(p,1)$ strings attached to the D3 brane. Similarly, we
construct  $(p,q)$ fluxons,  infinite $(p,q)$-strings piercing the
D3 brane. We match the tension of the gauge theory strings with
that of $(p,q)$-strings.

Section 8 contains some concluding remarks.

\newsec{Noncommutative Gauge Theory}

\subsec{Notations and setup}

Let us briefly review the framework of \nc gauge theory and
establish notation. Consider   space-time   with coordinates
$x^i$, $i=1, \ldots, d$ which obey the following commutation
relations: \eqn\cmrl{[x^i, x^j] = i {\t}^{ij}\ , } where
${\t}^{ij}$ is a constant asymmetric matrix of  rank $2r \leq d$.
By noncommutative space-time we mean the algebra ${\CA}_{\t}$
generated by the $x^i$ satisfying \cmrl , together with some extra
conditions on the allowed expressions of the $x^i$. The elements
of ${\CA}_{\t}$ can be identified with ordinary functions on ${\bf
R}^d$, with the product of two functions $f$ and $g$ given by the
Moyal formula (or star product):
\eqn\myl{ f \star g \, (x)=
{\exp} \left[ {i \over 2} {\t}^{ij} { {\p} \over {\p x_{1}^{i} }}
{{\p}\over {\p x_{2}^{j}}} \right] f(x_{1}) g (x_{2}) \vert_{x_{1}
= x_{2} = x}\ .}
For  plane waves:
\eqn\mylw{ e^{i {\vec p}_1
{\cdot } {\vec x} } \star e^{i {\vec p}_2 \cdot {\vec x}} = e^{-
{i\over 2} {\vec p}_1 \times {\vec p}_2}\quad e^{i ({\vec p}_1 +
{\vec p}_2) \cdot {\vec x}}\ , }
where
\eqn\vpr{{\vec p}_1 \times
{\vec p}_2 = {\t}^{ij} p_{1 i} p_{2 j} = - {\vec p}_2 \times {\vec
p}_1 \ .}
  We shall restrict our attention to the case of
${\CN}=4$ four dimensional $U(1)$ super-Yang-Mills theory on a \nc
space-time, with the noncommutativity parameter ${\t}^{\m\n}$
being space-like. One can then choose coordinates  so that
\eqn\cmt{[x^1, x^2] = -i {\t}, \quad [x^3, \cdot ] = [t, \cdot ] =
0 \ .}

\noindent The Lagrangian of a field theory involves derivatives.
The derivative ${\p}_i$ is the infinitesimal automorphism of the
algebra \cmrl: \eqn\auto{x^i \to x^i + {\ve}^i,} where ${\ve}^i$
is a $c$-number. For the algebra \cmrl\ this automorphism is
internal: \eqn\intrn{{\p}_i {\Psi} = i {\t}_{ij} [ {\Psi}, x^j],}
where ${\t}_{ij}$ is the inverse of ${\t}^{ij}$, namely
${\t}_{ij}{\t}^{jk}=\delta^k_i$. It is convenient to introduce the
operators: \eqn\osc{c = {1\over{\sqrt{2\t}}} \left( x^1 - i x^2
\right),\quad c^{\dagger} = {1\over{\sqrt{2\t}}} \left( x^1 + i
x^2 \right),} which obey: $$ [ c , c^{\dagger} ] = 1.$$
Note that
\eqn\drvt{{{\p}\over{{\p} x_{1}}} = {1\over{\sqrt{2{\t}}}} [ c -
c^{\dagger}, \cdot],\quad {{\p}\over{{\p} x_2}} =
{i\over{\sqrt{2\t}}} [ c + c^{\dagger}, \cdot ]\ .}
\ndt Since $c,
c^{\dagger}$ satisfy the commutation relations of the annihilation
and  creation operators we can identify functions $f(x_1,x_2)$
with functions of the $c,c^{\dagger}$ valued in the operators
acting in the standard Fock space ${\CH}$ of the creation and
annihilation operators:
\eqn\fsp{\eqalign{ {\CH} & =
\oplus_{n=0}^{\infty} \, {\bf C} \vert n \rangle\ , \cr
c^{\dagger} &
\vert n \rangle = \sqrt{n+1} \vert n+1 \rangle, \cr
c  & \vert n
\rangle = \sqrt{n} \vert n-1 \rangle\ ,\cr & {\hn} = c^{\dagger}c,
\quad {\hn} \vert n \rangle = n \vert n \rangle \ , \cr
& \qquad
\langle m \vert n \rangle = {\d}_{mn} \ .\cr}}
Since we shall be
dealing with a scale invariant theories in which the only scale
is $\t$ we shall set $2\t=1$. When desired, $\t$ can be
introduced back simply by rescaling the coordinates, $x_i \to
x_i/\sqrt{2\t }$, for $i=1,2$.

The procedure that maps ordinary commutative functions onto
operators in the Fock space is called Weyl ordering and is defined
by: \eqn\wlor{f(x)=f\left( z=x^1- ix^2, \bar z=x^1+ix^2\right)
\mapsto  {\hat f(c, c^\dagger)} =  \int f(x) \, {{{\rm d}^{2} x
{\rm d}^{2} p }\over{(2{\pi})^{2}}} \, \,  e^{ i \left[ {\bar p}_a
\left(  c  - z \right) + p_{a} \left( {c^\dagger}_a - {\bar z}
\right) \right] } \ . } It is easy to see that \eqn\product{  {\rm
if } \quad f \mapsto \hat f, \quad g \mapsto \hat g \quad {\rm
then }\quad f\star g \mapsto \hat f \hat g \ .} \ndt A useful
formula is for the matrix elements of $\hat f$ in the coherent
state basis \eqn\coherent{\langle {\xi} \vert {\hat f} \vert
{\eta} \rangle = \int f\left(z, {\bar z} \right) {{d z \, d
{\zb}}\over{(2{\pi}i)^{2}}} \, e^{\xi \cdot \eta - 2 ( {\xi} -
{\zb} ) \cdot ( {\eta} - z )}\ , } where $\langle {\xi} \vert$ and
$\vert {\eta} \rangle$ are coherent states: $\langle {\xi} \vert =
\langle {\bf 0} \vert {\exp} \left( {\xi} c^\dagger \right) ,
\qquad \vert {\eta} \rangle = {\exp} \left( {\eta} c \right) \vert
{\bf 0} \rangle .$ Translations in the Hilbert space are generated
by $\hp_i $, where \eqn\hatp{ \hp_1=  (c-c^\dagger)=-2i x_2, \quad
\hp_2=i (c+c^\dagger)=2i x_1  \ .} Thus if $ f(x) \mapsto \hat f$,
then $ f(x +a) \mapsto \exp(a\cdot \hp)\hat f\exp(- a\cdot \hp)$ .

\subsec{Gauge theory}
The covariant derivative of a $U(1)$ gauge
field is then represented as the operator:
\eqn\covder{\eqalign{&
D_0=\p_0 +A_0,\quad D_3=\p_3 +A_3  , \cr &  D = {\half} \left( D_1 +i
D_2 \right) = -c^\dagger+A , \quad \bar D= {\half} \left( D_1 -i D_2
\right) =c + \bar A, }}
where $A_\mu$ are the anti-Hermitian
components of the gauge field and $$ A =  {\half} \left( A_1 + i
A_2 \right), \quad \bar A = - A^{\dagger} = {\half} \left( A_1 - i
A_2 \right) .$$ Under a gauge transformation $$D \to UD U^\dagger,
\ \ \bar D \to U \bar D U^\dagger; \quad U^\dagger U=U U^\dagger=1
\ .$$ The anti-Hermitian field strength is $F_{\mu\nu} = [D_\mu,
D_\nu] - i{\t}_{\m\n}$.

The action for the  ${\CN}=4 $ supersymmetric \nc $U(1)$ gauge
theory is given  by ($a=1...6)$: \eqn\glagr{{\CL}(A) = -
{{2{\pi}{\t}}\over{g^2}} \int {\rm d}t {\rm d}x_3 \, {\Tr} \left[
{1\over 4}F_{\mu\nu}F^{\mu\nu} +{1\over 2} D_\m \Phi_a D^\m \Phi_a
+ {1\over 4}[\Phi_{a}, \Phi_ b]^2\right] + {\rm fermions} \ ,}
where the trace ${\Tr}$ over the Fock space states is equivalent
to integration over the noncommuting coordinates $x_1$ and $x_2$.

\subsec{Topological charges in \nc gauge theory}

Just as in an  ordinary gauge theory,  \nc gauge theory has
topological charges, e.g. magnetic fluxes or instanton numbers.
These can be defined via integrals of the characteristic forms:
\eqn\tpch{\int {\Tr} e^{{F \over{2\pi i}}}\ ,}
where  now the
integration over the \nc space-time $\int$ can be included in the
definition of the trace ${\Tr}$, with all the factors $2{\pi}
{\t}$ understood. In the commutative case there is an alternative
definition, which involves patching the space-time with  open
domains, glueing functions etc. In the \nc case such a definition
is lacking simply because one has to work (in some sense) globally
over the noncommutative part of the space-time. However, for \nc
${\bf R}^{2}, {\bf R}^4$ there are ``asymptotic'' techniques for
calculations of the topological charge.

Let us discuss \nc ${\bf R}^2$ for simplicity. Recall that we view
the components $A_1, A_2$ of the gauge field on the \nc ${\bf
R}^{2}$ as operators in the Fock space ${\CH}$. Suppose we are
looking for   gauge field configurations with finite energy
density when integrated over ${\rm d}x^1 {\rm d}x^2$. Then, as $x_1^2 +
x_2^2 \to \infty$ (in other words, when looking at the matrix
elements of the operators between the states of high occupation
numbers) the gauge fields must approach a pure gauge:
\eqn\asm{A_{\m} \to U^{\dagger} {\p}_{\m} U \ ,}
where $U \in
U({\CH})$ is a unitary operator in ${\CH}$. More precisely, since
we understand the limit in \asm\ to mean:
\eqn\lmt{\langle n \vert
A_{\m} \vert {\nb} \rangle \to \langle n \vert U^{\dagger}
{\p}_{\m} U \vert {\nb} \rangle, \quad {\rm as }\quad n, {\nb} \to
\infty\ , }
we only require that $U$ is well-defined and unitary on
the subspace of ${\CH}$ which contains the states $\vert n
\rangle, \vert {\nb} \rangle$ with sufficiently high $n, {\nb}$.
We can then continue $U$ on the whole of ${\CH}$ but it will,
generically cease to be unitary on the whole of ${\CH}$. The
measure of the non-unitarity of $U$ is its index: $$ {\rm Ind} U =
{\rm dim} {\rm Ker} U - {\rm dim} {\rm Ker} U^{\dagger} \ .  $$ If
$U$ can be deformed to the unitary operator then certainly ${\rm
Ind} U = 0$. If, on the other hand, ${\rm Ind} U \neq 0$ then $U$
cannot be deformed into a unitary operator. We could just as well
take the difference of the dimensions of the kernels of the
Hermitian operators to define the unitarity obstruction:
 $${\CI}_{U} = {\rm dim}{\rm Ker} U^{\dagger}U
- {\rm dim}{\rm Ker} UU^{\dagger}. $$ The solutions which we shall
discuss below will have ${\CI}_{U} \neq 0$.

\newsec{Monopoles and Strings}

\subsec{BPS solitons}

In our previous paper, \grossnek\ we discussed the \nc
generalization of Nahm's equations that describe BPS solitons of
gauge theory. We considered classical, static, solutions of the
theory given by \glagr . To construct these solutions we set
$A_0=0$ and chose $\Phi_a=\delta_{a1} \Phi$. Of course, given any
particular solution for $\Phi$ we can write the general solution
as $\Phi_a= \hat {\bf n} \Phi$, where $\hat {\bf n}$ lies on
$S^5$. For this choice the BPS equations, that minimize the
energy, are \eqn\BPS{\left[ D_i, \Phi\right]= \pm B_i,\quad
B_i=\ha \e_{ijk}F_{jk} \ .} We were able to find explicit
solutions of these equations, by using the \nc Nahm equations. The
solutions were given by the following expressions:
\eqn\hgsn{{\Phi} = {\Phi} ( {\hn} ) =
{{{\z}({\hn})}\over{{\z}({\hn}-1)}} -
{{{\z}({\hn}+1)}\over{{\z}({\hn})}}\ , }
\eqn\Aold{A_3=0; \quad
A=c^{\dagger}\left( 1 - {{\xi}({\hn}) \over {\xi}({\hn}+1)}
\right) \ , } where the functions $\xi( {\hn}), \ \z ({\hn})$ are
given in \grossnek. We shall not need here the explicit form of
these functions.

The above soliton looks very simple  far away from the origin,
where it reduces to fields that describe a semi-infinite string
along the positive $x_3$ axis plus a magnetic monopole at the
origin. In particular, $$\Phi \to -2x_3 P_0;\quad B_3\to 2P_0\ ;
{\rm  \ for \ large \ \ } x_3>0 \ ,$$ where $P_0= \vert 0 \rangle
\langle  0\vert $ \ is the projection operator in the Fock space
${\rm \cal H}$ which projects onto the vacuum state. When
translated back to ordinary position space the operator $P_0$
becomes $\exp\left[-{{x_1^2+x_2^2}\over{\t}}\right]$.     Thus
this soliton looks like a magnetic monopole at the origin that is
attached to a semi-infinite string along the positive $x_3$ axis.
In \grossnek\ we argued that this soliton corresponds precisely to
a D1 string attached to a D3 brane at the origin. We
  calculated the tension (=energy per unit
length) of the string and found that
  it is given by\eqn\tension{T= {2\pi\over g^2\t}   \  , }in complete
agreement with the tension of a D1 string in the bulk, correctly
scaled in the Seiberg-Witten decoupling limit. The D1 string,
tilted by the background B-field, forms an angle ${\psi}$,
${\tan}{\psi} = {{2\pi {\ap}}\over{\t}}$ with the D3 brane.
Although this angle goes to zero in geometrical units in the
Seiberg-Witten $\ap \to 0$ limit, in the gauge theory this angle
can be observed by the asymptotic slope of the Higgs field, since $\ap$ is
hidden when we replace the coordinates tranverse to the D3 brane
(which have dimensions of length) by the scalar  fields on the
brane (with dimensions of mass).

\subsec{An infinite string}

The solution \hgsn, \Aold,  discussed above
describes a semi-infinite string,
attached to a monopole--a reflection of a D1 string attached to a D3 brane.
There are other BPS string configurations,
whose decoupling limit in the
presence of a background B-field
might also  give rise to noncommutative
gauge theory solitons. We might
consider an infinite D1 string
that pierces the D3 brane. This is certainly BPS. The semi-infinite
D1 string attached to the D3 brane can be
regarded as the limit
of a D1 string stretched between two D3
branes, in the limit where the
separation between the D3 branes goes to
infinity. Before taking the
limit, this corresponds,
to a monopole in the $U(2)$ gauge theory,
that describes the low energy dynamics of
the D3 branes, broken to
$U(1)$ by   separating
 the branes. The distance between the
branes is proportional to the
vacuum expectation value
of the Higgs field  and, when this goes to
infinity, only the $U(1)$ gauge
degrees of freedom, and the massless Higgs field survive.

To obtain the piercing string we could start with 3 D3 branes and
break  the $SU(3)$ gauge theory to $U(1) \times U(1) \times U(1)$
by separating all the branes. We can then stretch one D1 string
from the middle D3 brane up (in the $\Phi$ direction) and another
D1 string down. In the limit of infinite separation,  and in the
presence of a background B-field, this will describe two
semi-infinite strings, in the $+x_3$ and in the $-x_3$ directions.
If we bring their origins together at the same point on the D3
brane we will obtain an infinite, piercing string---the fluxon.
The monopole and the   and anti-monopole at the intersectiuon
point will annihilate and we should be left only with the flux
tube.

We can obtain the piercing string solution by  manipulating the
solution described in \grossnek. Consider the Higgs field
$\Phi(x_3)$. Far away from the origin it approaches $-2x_3 P_0$,
up to terms that fall off as $1/x_3$. If we consider a solution
translated in the $x_3$ direction by an amount ${\d}$, in the
limit as ${\d} \to \infty$, then for any finite $x_3$,
$\Phi=-2(x_3+{\d})P_0 +O(1/{\d})$. It is easy to verify that the
components $A, \bar A$ are given, in this limit by, \eqn\trns{A =
c^{\dagger} \left( 1 - \sqrt{{\hn}\over{{\hn} +1 }} \right), \quad
\bar A = \left( \sqrt{{\hn}\over{{\hn} + 1}} - 1\right) c\ ,}
where $\hat n$ is the number operator: ${\hn} = c^{\dagger}c$.
Equivalently, \eqn\trnst{A=\sum_{n=0}^\infty
\left(\sqrt{n+1}-\sqrt{n}\right) \vert n+1\rangle \langle n \vert
\ . } Finally, we can drop the $x_0P_0$ term in $\Phi$, since the
equations of motion for $\Phi$ only involve derivatives with
respect to $x_3$. Thus, the above gauge fields and
\eqn\phisol{\Phi=-2x_3P_0 \ , } yield a BPS solution of the static
gauge theory.

It is easy to check that this is indeed a BPS soliton. For the
above configuration, $A_3=0$ and   $A_1 $and $A_2$ are independent
of $x_3$. Thus it follows that the only nonvanishing component of
the magnetic field is $B_3$  since $B_i
\propto\epsilon_{ij}\partial_3A_j, \quad i,j=1,2,$ and\eqn\Bthr{
B_3 = 2\left( [ \bar D, D]+1  \right),} where \eqn\Dis{D=
-c^\dagger\sqrt{ {\hn} \over{{\hn} +1}}, \ \bar D=\sqrt{ {\hn}
\over{{\hn} +1}}c \ \Rightarrow \  -\bar D D= \hn, \quad -\ D \bar
D = \hn -1 +P_0\ .} This is consistent with the BPS equations
since $\Phi  \propto P_0$ and thus \eqn\phoeq{\hn P_0 = P_0\hn= c
P_0 =P_0 c^\dagger =0 \Rightarrow  [D,\Phi] =
 [\bar D, \Phi]=0 .}
It is instructive to check the equations \phoeq\ in the coordinate
space, for example: $$c \, P_0 \mapsto (x_1 - i x_2 ) \star \exp
\left( - 2 \left(x_1^2 + x_2^2 \right)\right)  = 0     . $$
Finally, we have \eqn\Ds{ B_3 = 2 P_0 = -\p_3 \Phi \ .} This
solution clearly describes an infinite flux tube of magnetic field
along the $x_3$-axis, localized in the \nc plane, with a linear
$\Phi$ field along the tube that corresponds to the extension of
the tube into a direction transverse to the D3 brane.
Consequently, we shall call this soliton a {\it
fluxon}\footnote{$^{\dagger}$}{A similar solution was constructed in
\alexios, Eq.(31)}.

The fluxon solution for the gauge field is almost a pure gauge. We
can write the covariant derivatives, $\bar D$ and $D$, as
\eqn\bard{\eqalign{\qquad & \bar D  = S c S^{\dagger}, \ \ D = - S
c^{\dagger} S^{\dagger} ,  \cr \quad {\rm where} \quad S& =
c^\dagger {1\over \sqrt{\hn+1}}= \sum_{n=0}^\infty\vert
n+1\rangle\langle n \vert\ , \quad S \vert n\rangle =\vert
n+1\rangle \ ,\cr S^\dagger &={1\over \sqrt{\hn+1}}c
=\sum_{n=0}^\infty\vert n \rangle\langle n+1\vert  \ , \quad
S^\dagger \vert n\rangle =\vert n-1\rangle, \quad n >0,  \
S^{\dagger} \vert 0 \rangle = 0 .  \cr }   } If $S$ was unitary
this would be a pure gauge, but \eqn\uunit{ S^\dagger S=1, \quad
SS^\dagger= 1- P_0\ .} Thus $S$ is unitary in the subspace picked
out by $1-P_0$, and only fails to be unitary in the vacuum state.
In terms of the indices discusses above, the index of $S^\dagger$ or of $SS^\dagger$ is:
$$ {\rm Ind}S^\dagger ={\CI}_{S^\dagger}=1 .$$

\noindent It is also interesting to consider  the commutative
limit ${\t} \to 0$ of the fluxon solution. The vacuum projector
corresponds to the Gaussian packet in the noncommutative plane:
$$P_0 \to 2\exp\left[-{{x_1^2+x_2^2}\over{\t}}\right] \to  2\pi\t
{\d}(x_1) {\d} (x_2)\ {\rm as} \  \t \to 0\ . $$ Thus in this
limit the fields are given, in coordinate space  by:
\eqn\clslmt{{\Phi} = - 2{\pi} {\d}(x_1) {\d} (x_2) x_3, \quad A =
i {{x_1 \, dx_2 - x_2 \, dx_1}\over{x_1^2 + x_2^2}}\ ,} and the
magnetic field  is
\eqn\crvt{B=dA = 2\pi i {\d}(x_1) {\d}(x_2)\ ,}
 which clearly satisfies:
\eqn\bgmnl{dA + i \star d{\Phi} = 0 \ .}
Thus in the commutative theory we have a singular solenoid
extending along the $x_3$ axis.
\noindent The fluxon has a magnetic charge, defined as:
\eqn\mgch{Q_{m} = {1\over{2{\pi}i}} \int F_{12} \, {\rm d}x^1
\wedge {\rm d}x^2 = -i {\t} {\Tr} F_{12} = + 1\ . }

\subsec{Higher charge fluxons}

It turns out that a very simple modification of \bard\ produces
flux tubes of higher magnetic charge. The idea is to use the fact
that a subspace of the Fock space ${\CH}$, orthogonal to
a finite-dimensional subspace,
is  isomorphic to ${\CH}$ as a Hilbert space. Let $S_{N}$ be
a unitary isomorphism
\eqn\unis{S_{N} : {\CH} \to {\CH}_{N}, \quad
{\rm dim} \left({\CH}/{\CH}_{N}\right) = N\ .}
We assume that
\eqn\splt{{\CH} =
V_{N} \oplus {\CH}_{N}, \quad {\rm dim}V_{N} = N, \quad V_{N}
\perp {\CH}_{N}\ ,}
where $\perp$ means orthogonality in the
sense of the $\langle \vert \rangle$ hermitian inner product on
$\CH$. We have:
\eqn\prpt{S_{N}^{\dagger} S_{N} = 1, \quad
S_{N}S_{N}^{\dagger} = 1 - P_{N}\ ,}
where $P_{N}$ is the orthogonal
projection onto $V_{N}$.

\noindent Then the $N$-fluxon is given by:
\eqn\nflx{\eqalign{& D
= - S_{N} \, c^{\dagger} \, S^{\dagger}_{N}, \quad {\bar D} =
S_{N} \, c \, S_{N}^{\dagger}\ , \cr & {\Phi} = - 2x_3 P_{N} \ . \cr}}
\noindent By a unitary gauge transformation we can always bring
$S_{N}$ to the following form:
\eqn\shft{S_{N}=S^N, \quad S_{N} \vert n \rangle =
\vert n + N \rangle, \quad P_{N} = \sum_{m=0}^{N-1} \vert m
\rangle \langle m \vert \ .}
\ndt The magnetic charge (per unit length) of the $N$-fluxon is clearly
$N$:
\eqn\mgchn{Q_{m} = {1\over{2\pi i}} \int F_{12} \, {\rm d}x^1
\wedge  {\rm d} x^2 = N\ ,}
which also equals the index of $S_N^\dagger$:
 $$ {\rm Ind}S_N^\dagger ={\CI}_{S_N^\dagger}=1 .$$

The N-fluxon exhibited above breaks the $U(\infty)$ (more
precisely, $U({\CH})$) gauge symmetry \eqn\unitary{D \to U D
U^\dagger, \ \ \bar D \to U \bar D U^\dagger, \, \ \Phi\to U \Phi
U^\dagger, \quad U^\dagger U=U U^\dagger=1 , \quad U\in U( \CH ) \
.} down to $U(N)$, where the $U(N)$ acts within $V_N$. However, we
can break this symmetry further, all the way down to $U(1)\times
U(1)\times \dots U(1)$, and separate the N-fluxons by shifting:
\eqn\higss{\Phi \to -2x_3 P_N + P_N D_N P_N  \ , } where $D_N$ is
a constant diagonal $N\times N$ matrix, with eigenvalues $d_i,\
i=1\dots N$. This will clearly solve the BPS equations and
represent N-fluxons intersecting the D3 brane at positions $x_3
=d_i/2$. If all the $d_i$ are different then $U({\infty})$ will be
broken down to $U(1)^N$.

Also notice that, as expected,  the ${\CN}=4$ theory has solutions
describing D1 strings separated in all six directions:
\eqn\higgsd{{\bf\Phi} = -2x_3 {\hat{\bf n}} P_N + P_N {\bf D}_{N}
P_{N}\ ,} where ${\bf D}_{N}$ is a sextet (transforming in the
representation ${\bf 6}$ of the $SO(6)$ R-symmetry group) of
diagonal $N \times N$ matrices\footnote{$^{\spadesuit}$}{The
complete solution, which depends on $8N$ moduli, together with the
proof of its uniqueness, will be presented elsewhere \grossneki}.

\subsec{Properties of fluxons}
The energy density of the $N$-fluxon is
\eqn\engden{ {\CE}= {1\over 2g^2} \left( {\vec B}^2 +  4 [D,
{\Phi}] [ {\bar D}, {\Phi} ] + \left( {\p}_3 {\Phi} \right)^2
\right) = {4\over g^2}P_N \ , }
and thus the gauge invariant
energy per unit ($x_3$) length, the tension, is (restoring $\t$)
\eqn\ten{T ={{2 \pi N}\over g^2 \t}\ . }

The 1-fluxon  is not translationally invariant. It corresponds to
a D1 string that pierces the D3 brane at a specific point. Of
course we can obtain other solutions by translating $x\to x+a $.
This is obvious in the case of translations along the $x_3$ axis
and $\p_3 \Phi= -2P_0$ will be a normalizable (per unit length)
zero mode of the soliton ($\p_3 A=0$). Similarly, in the case of
the separated N-fluxon solution \higss\ there are N translational
zero modes that shift the $d_i$.

Translations of the \nc coordinates are
subtler. Translations of operators in the \nc\ directions are
generated by the operators $\hp$, defined in \hatp . Thus we can
translate our solution  by an amount $a=(a_1,a_2)$ by performing
$$\Phi \to  \exp(a\cdot \hp)\Phi \exp(-a\cdot \hp), \quad A \to
\exp(a\cdot \hp)A \exp(-a\cdot \hp) \ .$$ This is a gauge
transformation of the Higgs field, $\Phi \to U(a) \Phi
U^\dagger(a)$, where
\eqn\Utrans{U(a)=\exp(-a\cdot
\hp)=\exp(ia_i\t^{-1}_{ij}\hat x_j ) \ .} Acting on the gauge
field this transformation yields \eqn\gaugetr{ A \to U(a) A
U^\dagger(a) = \left[ U(a) (A-c^\dagger) U^\dagger(a)
+c^\dagger\right] + U(a)[ c^\dagger, U^\dagger(a)]\equiv \delta_1
A +\delta_2 A \ . }
The first term, $\delta_1 A$, in \gaugetr\ is
a gauge transformation and the second term, $ \delta_2 A$, is a
constant, c-number,  shift of the gauge field. \eqn\delA{ \delta_2 A = U(a)[
c^\dagger, U^\dagger(a)] =-(a^1+i a^2)\ . }  Both of these, gauge
transformations and constant shifts of the gauge field, are
symmetries of the action. What is unusual about \nc gauge theories
is that {\it translations in the \nc directions are equivalent to
a combination of a gauge transformation and a constant shift of
the gauge field.} This explains why in \nc gauge theories there do
not exist local gauge invariant observables, since by a gauge
transformation we can effect a spatial translation! This is
analogous to the situation in general relativity, where
translations are also equivalent to gauge transformations (general
coordinate transformations) and one cannot construct local gauge
invariant observables. The fact that spatial translations are
equivalent to gauge transformations (up to global symmetry
transformations) is one of the most interesting features of \nc
gauge theories. These theories are thus toy models of general
relativity---the only other theory that shares this property.

The gauge transformation that corresponds to a translation is a
{\it large gauge transformation} that does not approach the
identity at spatial infinity; namely $U\sim 1 + i\lambda$, where
$\lambda= -i a^k {\t}^{-1}_{kl}x^l$. Thus these gauge
transformations are analogous to the large gauge transformations
that take us from one winding number vacuum to another in
non-abelian gauge theory.

When the string is quantized we will have to introduce collective
coordinates for these zero modes and construct wave packets with
definite momenta, much as we construct $\vartheta$-vacua in
non-abelian gauge theory or states of definite momentum in general
relativity.

\newsec{Fluctuations of the string(s)}

We shall now discuss the fluctuations of the theory about the
fluxon solution.  We expect to find a
variety of fluctuations when we linearize the theory about the
fluxon. First, we certainly expect to find a continuous spectrum
of photons and scalars (plus fermionic partners), which look
like the modes of the theory in the vacuum, far away from the
soliton. From the point of view of the string theory, these modes
correspond to light fundamental strings attached to the D3 brane.
We will therefore refer to them as 3-3 modes.

In addition we might expect to find the fluctuations of the string
itself. Since, we argue, the \nc gauge theory soliton represents a
D1 string, we might expect to find the modes of a supersymmetric
D1 string propagating in 10 dimensions. The $N$-fluxon is a gauge
theory soliton that represents $N$ D1 strings. The fluctuations of
these strings, whose tension is of order $1/g^2\t$, cannot be seen
in a small $g^2$ expansion about the string. In other words, the
massive D1 string states have energies of order $1/g^2$, and to
see these we would require nonperturbative techniques. However,
there are other modes of the D1 strings that are visable in the
semiclassical domain of the gauge field theory---namely the modes
that arise from fundamental open strings attached to the D1
strings. In the $\ap\to 0$  limit we are taking these modes should
be described by a supersymmetric  $U(N)$ Yang-Mills theory living
on the 2 dimensional world sheet of the D1 strings. If all the
fluxons are at the same point the $U(N)$ gauge symmetry should be
unbroken. If they are separated, as in \higss,\higgsd, than the
gauge theory will be in the broken symmetry Higgs phase.
  We indeed find such a spectrum of
small fluctuations. We refer to these as 1-1 modes.

Finally, there are some extra stringy modes. Given a D1 string
that is attached to a D3 brane, one can attach a fundamental
string to the D1 at one end and to the D3 brane at the other.
These too should be reflected in the gauge theory in the presence
of our D1 soliton. In the decoupling limit we are considering, the
only modes of these fundamental strings that could survive would
have to be short and thus localized around the point ($\vec x=0$)
where the D-branes intersect. Indeed, we find such   modes,
localized about the origin, with discrete energies. We shall refer
to these as 1-3 modes.

Let us restrict our attention to the $N=1$ fluxon. We shall
discuss the generalization to the case of the $N$ fluxon later.
\subsec{Expansion of the action} It is
convenient to view the ${\CN}=4$ supersymmetric Yang-Mills theory
as a dimensionally reduced ten dimensional SYM theory. We shall
use the indices ${\m}, {\n}, \ldots = 0,1,2,3$ to denote four
dimensional quantities, the indices $i,j, \ldots= 1,2,3$ to denote
three dimensional quantities, and the indices $A,B, \ldots = 0, 1,
\ldots, 9$ to denote ten dimensional ones. The scalar fields can be
regarded as the extra six components of the ten-dimensional gauge
field, \eqn\phifour{\Phi_a =iA_{3+a}, \quad a = 1, \ldots 6, \quad
D_B = [ A_B, \cdot ], \quad B = 4, \ldots, 9\ . } They are Hermitian
since our gauge fields are anti-Hermitian. The Lagrangian  can
then be written as \eqn\lagten{\CL  = - {1\over 4g^2}  \left(
[D_A, D_B]^2 + {\bar\l}_{\dot \a} {\Gamma}_{B}^{\dot \a \, \a} [
D_{B}, {\l}_{\a}] \right) \ .} Here we have included the fermionic
partners, ${\l}_{\a}$, of the \CN=4 supermultiplet. We expand the
SYM action   about  the soliton, $A^0_B$, for which the only
nonvanishing components are for $B=1,2,3,4$. Expanding
\eqn\expan{A_B= A^0_B + ga_B\ ,} and fixing the    gauge by
imposing the condition \eqn\gcn{[D^0_{B},a_{B}] = 0} on the
fluctuations, we obtain:
\eqn\act{\eqalign{\CL = & \CL_0 + \CL_2 +
\CL_3 + \CL_4\ ,\cr
& \CL_0 = {{2\pi }\over{g^2 {\t}}}P_0 \ ,\cr
& \CL_2= - {1\over{2}} \left( (D^0_{B} a_{C})^2 + 2F^0_{BC} [ a_{B},
a_{C}] + {\bar\l} {\CD^0} {\l} \right) \ , \cr
& \CL_3 = -g \left( D^0_{B} a_{C} [ a_{B}, a_{C}] + {\bar\l} {\Gamma}_{A} [
a_{A},
{\l}] \right) \ , \cr
& \CL_{4} = - {g^2\over 4} [ a_{A}, a_{B}]^2 \ .\cr}}

The only nonvanishing components of the field strength in the
background given by  \trns, \phisol\ are:
\eqn\fs{F^0_{12} = - 2i P_0, \quad
F^0_{34}= 2i P_0\ . }
The only nonvanishing components of the
covariant derivative are
$D^0_A, \ A=0,1,2,3,4$ .

\subsec{The linearized action}

To linear order ($g^0$) we have simply a bunch of free gauge
fields, scalars and fermions. We wish to  diagonalize the
quadratic form $- {\half} a_{A} {\Delta} a_{A} + {\bar\l}_{\dot\a}
{\CD}_{\a \dot \a} {\l}_{\a}$ that appears in $S_2$. The bosonic
operator $\Delta$, corresponding to $S_2$, is: \eqn\opr{-{\Delta}
a_{C} \equiv -[D^0_{B},[D^0_{B}, a_{C}]] + 2 [F^0_{CB}, a_{B}]\ .}
Each component of the gauge field $a_{B}$ is an $(t,
x^3)$-dependent operator in the Fock space $\CH$. The single
fluxon corresponds to the splitting of the Fock space into the sum
of a one-dimensional subspace $V_1 = {\bf C} \vert 0 \rangle$ and
the orthogonal complement ${\CH}_{1}$: $$ {\CH} = V_{1} \oplus
{\CH}_{1}. $$ Accordingly, the space ${\rm End}({\CH})$ of the
operators in the Fock space  splits as a direct sum of subspaces:
\eqn\splt{{\rm End}{\CH} = V_{11} \oplus V_{13} \oplus V_{31}
\oplus V_{33}\ , }
where
\eqn\sbsp{V_{11} = V_{1} \otimes V_{1},
\quad V_{13} = V_{1} {\otimes} {\CH}_{1}, \quad V_{31} = {\CH}_{1}
\otimes V_{1}, \quad V_{33} = {\CH}_{1} \hat\otimes {\CH}_{1}\ .}
The operator $\Delta$ preserves each of these subspaces. We shall see
that these subspaces contain the 1-1, 1-3, (their conjugates 3-1)
and 3-3 modes respectively, hence their names.

Moreover, one can also split the space ${\bf R}^{10} = {\bf R}^2
\oplus {\bf R}^2 \oplus {\bf R}^{1,5}$ of the Lorentz indices $A$,
as follows: $$ (1,2) \oplus (3,4) \oplus (0 , 5 , \ldots , 9) .$$
For the $1,2$ subspace we introduce the components of the gauge
field $X = {\half} ( a_1 + i a_2 )$, for which
\eqn\onetwo{-{\Delta} X = - [D^0_A[D^0_A \ , X]] + 4[ P_0 , X] \
.} For the $3,4$ subspace we introduce the combination $Y= {\half}
(a_3 + i a_4)$, for which \eqn\threefour{ - {\Delta} Y= -
[D^0_A[D^0_A \ , Y]] - 4[P_0, Y] \ .} For the other $ B=5,\ldots,
9$ components of the gauge field the operator $\Delta$ acts in a
simpler fashion: \eqn\oth{- {\Delta} a_{B}= -[D^0_A[ D^0_A\ ,
a_{B}] \ .}

Actually, both $[P_0, \cdot]$ and
$[D^0[D^0 , \cdot]]$ preserve the
decomposition \splt, therefore the
classification of the
eigenstates according to their $11$, $13$,
$33$ types holds for any
$A = 0, \ldots, 9$.

Now let us discuss the operator \eqn\cvlp{\eqalign{ D^2 {\CO}
\equiv & \left[D^0_{A}, \left[ D^0_{A}, {\CO}\right]\right] = -
{\p}_t^2 \, {\CO} + {\p}_3^2 \, {\CO} + 2 [D , [\bar D , {\CO} ]]
+ 2 [ \bar D , [ D, {\CO} ]] - [{\Phi}, [ {\Phi}, {\CO} ]]\cr = &
- ({\p}_t^2 - {\p}^2_3) {\CO} \,  - 4x_3^2 \left[ P_0, \left[P_0,
{\CO} \right] \right] +  \cr & 2 \left[ -c^{\dagger}
\sqrt{{\hn}\over{{\hn}+1}}, \left[ \sqrt{{\hn}\over{{\hn}+1}} c,
{\CO} \right]\right] + 2 \left[\sqrt{{\hn}\over{{\hn}+1}} c,
\left[ - c^{\dagger} \sqrt{{\hn}\over{{\hn}+1}}, {\CO}
\right]\right]\, .\cr} }
When restricted to the $V_{ab}$ subspaces
in \splt\ it simplifies to:
\eqn\delv{\eqalign{ V_{11}: & \quad
-D^2 = {\p}_t^2 - {\p}^2_3 \ ,\cr
V_{13}, V_{31} : & \quad -D^2 =
{\p}_t^2 - {\p}_3^2 + 4x_3^2 + 2 ( 2{\hn}+1) \ ,\cr
V_{33}: & \quad -
D^2 = {\p}_t^2 -{\p}_3^2 - 2 ( {\pb}_{\zb} - z) ({\p}_{z} - {\zb})
- 2 ( {\p}_{z} - {\zb}) ({\pb}_{\zb} - z) \ .\cr}}
We have
introduced, in the above, the following conventions. For $V_{13}$
we define: ${\hn} \vert 0 \rangle \langle m \vert = m \vert 0
\rangle\langle m \vert$ and for $V_{31}$: ${\hn} \vert m \rangle
\langle 0 \vert = m \vert m \rangle \langle 0 \vert$. For $V_{33}$
we introduce the following representation. For an operator $A \in
V_{33}$ we define a function of two variables $z, {\zb}$ as
follows: \eqn\genf{ A ( z, {\zb}) = \sum_{k,l > 0} \langle k \vert
A \vert l \rangle {{z^{k-1}\, {\zb}^{l- 1}}\over{\sqrt{(k-1)!
 (l-1)!}}}\ .}
Then the operators $D , {\bar D} $ acting on $a \in V_{33}$ will
be represented as the operators \eqn\rerep{D \sim {\pb}_{\zb} - z,
\quad {\bar D}  \sim {\p}_{z} - {\zb}}acting on $A(z, {\zb})$.

\noindent The gauge condition \gcn\ also preserves the
decomposition \splt. For the individual subspaces the gauge
condition reads: \eqn\gcnv{\eqalign{V_{11}: & \quad - {\p}_t a_0 +
{\p}_3 a_3 = 0 \cr V_{13}, V_{31}: & \quad -{\p}_t a_0 + {\p}_{3}
( Y - {\bar Y} ) - [ {\Phi}, Y + {\bar Y} ] + 2 [ \bar D , X] - 2
[ D,  {\bar X}] = 0\cr V_{33} : & \quad -{\p}_t a_0 + {\p}_3 a_3 -
2 ( {\pb}_{\zb} - z) {{ \bar Y (z, {\zb})}} + 2( {\p}_{z} - {\zb})
Y (z, {\zb}) = 0 \ .\cr}}

\subsec{The spectrum} Now we can easily diagonalize $\Delta$. We
start with the $V_{33}$ subspace. In this space, for all values of
the index $A$, the operator ${\Delta}$ coincides with $D^2$ since
all components are orthogonal to $\vert 0\rangle$ and thus they
commute with $P_0$.  As such it has the following eigenvectors:
\eqn\sptrtr{a_{A} (t, x_3, z, {\zb}) \sim {\z}_{A} ({\o}, {\vec
k}) e^{i ( {\o}t - k_3 x_3 - k_{\zb} z - k_z {\zb} ) + z{\zb}}\ ,
} with   eigenvalues \eqn\eigtrtr{{\o}^2 - k_3^2 - 4 k_{z}k_{\zb}
\equiv {\o}^2 - {\vec k}^2\ .} The gauge condition \gcnv\ implies
that the polarization vector ${\z}_{A}({\vec k})$ must be
transverse: $ - {\o} {\z}_t + k_3 {\z}_3 + 2 k_{z} {\z}_{\zb}+ 2
k_{\zb} {\z}_{z} = 0$. In sum, this branch of the spectrum
describes photons, scalars (and as we shall see later, their
superpartners) propagating along the three-brane worldvolume. As
usual, for   on-shell quanta, ${\o}^2 = {\vec k}^2$ the gauge
condition has a residual gauge freedom, which can be used to set
${\z}_t = 0$.

The $V_{11}$ branch of the spectrum is also simple. All components
are proportional to $P_0$ and thus commute with $P_0$.
Consequently ${\Delta}$ coincides with $-{\p}_t^2 + {\p}_3^2$ for
all $A$ and its eigenvectors are: \eqn\sponon{a_{A} \sim {\z}_A
({\o}, k) e^{i ({\o} t - k x_3)}\ , } with eigenvalues ${\o}^2 -
k^2$ and the condition on the polarization is: $$- {\o} {\z}^0 + k
{\z}^3 = 0\ . $$ Again, for the on-shell quanta, ${\o} = \pm k$,
and one can gauge both ${\z}^0$ and ${\z}^3$ away, leaving eight
physical modes. These modes correspond to the ground states of the
open fundamental strings attached to the D1 string. Indeed, these
are the modes of a 1+1 dimensional ${\CN}=8$ supersymmetric $U(1)$
gauge theory living on the world sheet of the D1 string.

Finally, we need to solve for the $V_{13}, V_{31}$ branches of the
spectrum.  Let us look at $V_{13}$ spanned by the operators
${\CO}_m = \vert 0 \rangle \langle m \vert, \quad m
> 0$. The discussion of $V_{31}$ will be similar. The
spectrum of the operator ${\Delta}$ depends on the index $A$. All
eigenfunctions have the same form: \eqn\egfn{a_{A} (t, {\vec x})
\sim {\z}_{A} ( {\o}, n,m) \, H_{n} (x_3 ) e^{-x_3^2} \,
{\CO}_{m}\ .} Here $H_{n}(x_3)$ denotes the $n$'th normalized
Hermite polynomial: \eqn\herm{H_{n}(x) = \left( {2\over{\pi}}
\right)^{\scriptstyle{1\over 4}} {1\over{\sqrt{2^n \, n!}}}
e^{-x_3^2} {\p}_3^{n} e^{x_3^2} \  . }

\noindent The eigenvalues of $\Delta$ depend on $A$. There are
three cases:

\noindent \sssec{1,2} the eigenvectors of $\Delta$ are: $$ X(n,m)
\sim H_{n} (x_3) \, e^{i {\o} t -x_3^2} {\CO}_m, \quad m > 0, \, n
\geq 0 \ , $$ with   eigenvalue: $${\o}^2 - 4 ( n + m +1) .$$

\noindent
\sssec{3,4} the eigenvectors of $\Delta$
are: $$ Y (n,m)
\sim H_{n} (x_3) \, e^{i {\o} t -x_3^2}
{\CO}_m,\quad m > 0, \, n \geq 0\ , $$ with
eigenvalue: $$ {\o}^2 - 4 (n+m-1)\ .$$

\noindent
\sssec{5,6,7,8,9} the eigenvectors of
$\Delta$ are:
$$ Z (n,m) \sim H_{n} (x_3) \, e^{i {\o} t - x_3^2} {\CO}_m, \quad  Z = a_A,
\,\, A \neq
1,2,3,4, \quad m > 0, \, n \geq 0\ ,$$
with  eigenvalue: $$ {\o}^2 - 4 (n+m)\ .$$

\noindent
The gauge condition imposes a relation
between $a_0$, $X$ and $Y$.
As for the remaining five components,
$a_A$, $A > 4$, there is no constraint.

In this sector the spectrum is discrete and the states are
localized about $x_3=0$. This is expected for the 1-3 modes, which
we argued should come from fundamental strings, stretched between
the D1 string and the D3 brane. Notice that the $m =1, n=0$ branch
of the spectrum $3,4$ corresponds to a zero mode. We discuss its
meaning below. The $5,\ldots , 9$  and $1,2$ branches of the
spectrum have a mass gap.

\subsec{Normalization of the modes}

We have three branches of the spectrum of the bosonic fluctuations
around our solution. We have to normalize them properly.

The norm on the fluctuations comes from the natural metric on the
gauge fields:
\eqn\norm{\Vert a_{A} \Vert^2 = - \int {\rm d} t
{\rm d} x_3 \, {\Tr} \, a_{A}^2\ . }
With the gauge condition
$D_{B}^{0} a_{B} =0$, the norm on the gauge fixed fluctuations is
given by the same formula \norm.

The generic fluctuation can be decomposed as follows:
\eqn\dcmps{\eqalign{ a_{A} ({\vec x}, t) = & \int i {{d {\o} d
k}\over{(2\pi)^2}} \, {\z}_{A} ({\o}, k) e^{i ({\o} t - k x_3)}
\vert 0 \rangle\langle 0\vert \, + \cr & \int
i{{d{\o}}\over{2{\pi}}} \, \sum_{n \geq 0, m \geq 1}  {\z}_{A}
({\o}, n,m) H_{n} (x_3) e^{i{\o} t -x_3^2}  \vert 0 \rangle
\langle m \vert \, + \cr & \int i{{d{\o}}\over{2{\pi}}} \, \sum_{n
\geq 0, m \geq 1}  {\bar\z}_{A} (-{\o}, n,m) H_{n} (x_3) e^{i{\o}
t -x_3^2}  \vert m \rangle \langle 0 \vert \, + \cr
 & \int i {{d {\o} d^{3} {\vec k}}\over{(2\pi)^4}} \,  {\z}_{A}
({\o},  {\vec k} ) \, S \, e^{i ({\o} t - {\vec k} \cdot {\vec
x})} \, S^{\dagger} \ .\cr}}
The condition that $a_{A}$ is
anti-Hermitian translates into: \eqn\ahc{{\z}_{A} (- {\o}, -k )=
{\bar\z}_{A} ({\o}, k), \quad {\z}_{A} ({\o}, {\vec k}) =
{\bar\z}_{A} (- {\o}, - {\vec k})\ .}
 Now we substitute \dcmps\
into \norm\ and get: \eqn\nrmls{\eqalign{ & \qquad\qquad \Vert
a_{A} \Vert^2 = \cr & \int {{d{\o}}\over{2\pi}} \left( \int {{d
k}\over{2{\pi}}} \vert {\z}_{A} ({\o}, k) \vert^2 +
\sum_{\scriptscriptstyle{n \geq 0, m \geq 1}} {\bar\z}_{A} (-{\o},
n, m) {\z}_{A} ({\o}, n, m) + \int {{d^{3} {\vec
k}}\over{(2\pi)^3}} \vert {\z}_{A} ({\o}, {\vec k}) \vert^2
\right) , \cr}} where we used the formulae: \eqn\trcs{{\Tr} S B
S^{\dagger} = {\Tr} B, \quad {\Tr} S^{\dagger} B S =  {\Tr}  B -
\langle 0 \vert B \vert 0 \rangle, \quad {\Tr} e^{i ( {\kk}
c^{\dagger} + {\bar\kk} c )} = {\d}^{(2)}({\kk}) \ . }

Summarizing, the fluctuations are decomposed as in \dcmps\ and the
eigenvalues of the various terms in \dcmps\ are given by:
\eqn\eigv{\eqalign{V_{11}: \, & \, {\o}^2 - k^2 \ ,\cr V_{13},
V_{31}: \, & \, {\o}^2 - 4 ( m + n + (-1, 0 ,1))\ , \cr V_{33}: \,
& \, {\o}^2 - {\vec k}^2 \ , \cr}} from which we can read off
the propagators.

\subsec{Moduli and translational modes.} The spectrum of the
fluctuations computed above contains three types of zero modes: in
the $1-1$, $1- 3$ and $3-3$ sectors respectively. In the $1-1$
sector we should find the translational zero modes of the soliton.
The zero mode corresponding to translations in the $x_3$ direction
is simply $\p_3\Phi=- 2P_0$ , or more generally ${\z}^A ({\o}=0,
k=0) \, P_0$, eight for all eight transverse components of the
string. The translational modes in the \nc directions, as
discussed above, are infinitesimal gauge transformations.

Let us now look at the $1-3$ sector. Here we find a zero mode that
we shall interpret as a moduli of the solution space of soliton
strings. This zero mode is given by: \eqn\otzm{a_3 + {\varphi} =
{\eta} \vert 0 \rangle \langle 1 \vert \, e^{-x_3^2}\ ,} where
$\eta$ is a complex number, from which we get:
\eqn\phizm{{\varphi} = {\half} \left( {\eta} \vert 0 \rangle
\langle 1 \vert + {\bar\eta} \vert 1 \rangle\langle 0 \vert
\right) e^{-x_3^2}\ .} To understand this mode it is instructive
to look at the eigenvalues of the perturbed Higgs field:
\eqn\prthgs{{\Phi}_{\e} = {\Phi} + {\e} {\varphi} = - 2x_3 \,
\vert 0 \rangle \langle 0 \vert + {\e \over 2} \left( {\eta} \vert
0 \rangle \langle 1 \vert + {\bar\eta} \vert 1 \rangle\langle 0
\vert \right) e^{-x_3^2}\ .} Let us redefine ${\e}{\eta}/2 \to
{\eta}$. The matrix \prthgs\ is a $2 \times 2$ matrix which is
simple to diagonalize: the eigenvalues are the roots of the
quadratic equation: \eqn\egn{{\l} ( {\l} + 2x_3) = \vert \eta
\vert^2 e^{-2x_3^2} \ .} For $x_3 \to \pm \infty$ one of the
eigenvalues goes to zero, while the other behaves like $-2x_3$.
Since $\vert \eta \vert^2 e^{-2x_3^2} > 0 > - x_3^2$ the
eigenvalues never cross. It means that the spectral surface
associated with $\Phi$ contains two branches which look like two
semi-infinite strings, separated by the distance of  order $\vert
\eta \vert$. They become visible if we rewrite  \prthgs\ in
position space as (we choose $\eta=1$),
\eqn\possp{\Phi_\e(x_1,x_2,x_3)= \exp[-2(x_1^2+x_2^2)]\left[-2x_3
+2\e x_1 \exp(-x_3^2)\right]\ . } The interpretation of this zero
mode is clear---it corresponds to splitting the fluxon into two
semi- infinite strings of the type constructed in \grossnek. If we
were to continue to separate these strings they would approach,
for large separation in the $x_1$ direction, two semi-infinite
strings as constructed in \grossnek\ . It is not trivial to
construct the explicit solution for two semi-infinite strings
separated by a finite distance, since we would have to relax the
simplifying axially symmetric ansatz of \grossnek\ .

\newsec{Interactions}

The interactions of the  various modes
in the presence of the soliton are described by the nonlinear terms in the
Lagrangian,
$\CL_3$ and $\CL_4$.

The vertices of the gauge theory are organized in a manner which
mimics the disc amplitudes of the open string theory with the D1
or D3 boundary conditions. The cubic vertices correspond to the
three point correlation function, while the quartic vertices come
from the four point function (the latter also contains the
contribution of the tree diagram with two cubic vertices).

To proceed further we need to decompose the product $a_{A}a_{B}$:
\eqn\prdc{a_{A} a_{B} \, = \left( a_{A}a_{B} \right)^{11} \vert 0
\rangle \langle 0 \vert + \left(a_{A}a_{B}\right)^{13}_{m} \vert 0
\rangle\langle m \vert + \left( a_{A}a_{B}\right)^{31}_{m} \vert m
\rangle \langle 0 \vert + \left( a_{A}a_{B}\right)^{33}_{m{\mb}}
\vert m \rangle\langle {\mb} \vert \ .}
We shall omit summations over
 repeated indices or integrations over repeated momenta. We
also omit the factor $e^{i({\o} + {\o}^{\prime})t}$ in front of
everything, and we assume that  frequency of the $a_{A}$ mode is
${\o}$, while $a_{B}$ has the frequency ${\o}^{\prime}$ (the
frequencies are to be integrated over, of course). Finally, ${\vec
k} = (k, {\kk}, {\bar\kk})$, ${\vec k} \cdot {\vec x} = k x_3 +
{\kk} c^{\dagger} + {\bar\kk} c$.

We have:
\eqn\dcmpi{\eqalign{ & \left( a_{A}a_{B} \right)^{11}
 \, = \, \left( {\z}_{A} ({\o}, k) {\z}_{B} ({\o}^{\prime}, k^{\prime})
 e^{i (k+k^{\prime})x_3} +
 {\z}_{A} ({\o}, m,n) {\bar\z}_{B} ({\o}^{\prime}, m,l)
 H_{n}(x_3) H_{l}(x_3)e^{-2x_3^2} \right) , \cr
&  \left( a_{A}a_{B} \right)^{13}_{m} =  H_{n}(x_3)
 \exp{- (ik x_3  +  x_3^2)}\quad  \times \cr
 & \qquad \left(
 {\z}_{A}({\o},k) {\z}_{B} ({\o}^{\prime}, m,n)
 + {\z}_{A} ({\o}, {\mb}, n) {\z}_{B} ({\o}^{\prime}, {\vec k})
 \langle{\scriptstyle{{\mb}-1}}\vert e^{-i ({\kk} c^{\dagger} + {\bar\kk} c) }
 \vert{\scriptstyle{m-1}}\rangle
 \right) \ ,\cr
& \left( a_{A} a_{B} \right)^{31}_{m} = \overline{\left( a_{B}
a_{A}
 \right)^{13}_{m}}\ , \cr
&  \left( a_{A} a_{B} \right)^{33}_{m{\mb}} = {\bar\z}_{A} ({\o},
m, l) {\z}_{B} ({\o}^{\prime}, {\mb} , p) H_{l}(x_3) H_{p}(x_3)
e^{-2x_3^2} + \cr & \qquad\qquad {\z}_{A}({\o}, {\vec k}) {\z}_{B}
({\o}^{\prime}, {\vec k}^{\prime}) e^{- {i\over 2} {\vec k} \times
{\vec k}^{\prime}} \langle {\scriptstyle{m-1}} \vert e^{-i ({\vec
k} + {\vec k}^{\prime}) \cdot {\vec x}} \vert {\scriptstyle{{\mb}
-1}} \rangle\ . \cr}}

\subsec{The cubic interactions} Let us look at the cubic term
$\CL_3$ in the expansion of the Yang-Mills Lagrangian about the
fluxon solution.  It, of course,  contains the usual interactions
of the bulk photons and their superpartners. More interesting is
that it contains interactions of the modes from the $V_{11}$
sector, i.e. the fluctuations of the string, with the $V_{13},
V_{31}$ modes. In turn, these modes can annihilate into bulk
modes.

We shall also need the expressions for $D_{B}^{0}a_{A}$:
\eqn\drv{\eqalign{& D_{t}^{0}a_{A}({\o})  = i{\o} a_{A}({\o}),
\quad  D_{3}^{0} a_{A}^{11,33} = -i k a_{A}^{11,33}, \cr &
D_{3}^{0} a_{A}^{13} = {\z}_{A} ({\o}, m, n) \left( \sqrt{n}
H_{n-1} (x_3) - \sqrt{n+1} H_{n+1} (x_3) \right) e^{i {\o} t -
x_3^2} \vert 0 \rangle\langle m \vert \ ,\cr & D_{4}^{0}
a_{A}^{13} = - i{\z}_{A} ({\o}, m, n) \left( \sqrt{n} H_{n-1}
(x_3) + \sqrt{n+1} H_{n+1} (x_3) \right) e^{i{\o}t - x_3^2} \vert
0 \rangle\langle m \vert \ ,\cr & {\bar D}^{0} a_{A}^{13} =
{\z}_{A} ({\o}, m, n) H_{n} (x_3)  e^{i {\o} t - x_3^2}\quad
\sqrt{m-1} \vert 0 \rangle\langle m-1 \vert \ , \cr & D^{0}
a_{A}^{13} = - {\z}_{A} ( {\o}, m, n) H_{n}(x_3) e^{i{\o}t -x_3^2}
\quad \sqrt{m} \vert 0 \rangle \langle m+1 \vert \cr & D^{0}
a_{A}^{33} = -i {\bar\kk} {\z}_{A} ({\o}, {\vec k}) \, S \,
e^{i({\o}t - {\vec k}\cdot {\vec x})} \, S^{\dagger}\ , \cr}} as
well as  the overlap integrals/matrix elements:
\eqn\trvrt{\eqalign{& {\CV}_{\Vert}(k, n, {\nb}) =\int dx\,
e^{-ikx - 2x^2} H_{\nb} (x) H_{n} (x)\ , \cr
&{\CV}_{\perp} ({\kk}, n,
{\nb} ) = \langle {\nb} \vert e^{-i ({\bar\kk} c + {\kk}
c^{\dagger})} \vert n \rangle \ .\cr}}
\noindent The integrals
\trvrt\ are easy to evaluate using the oscillator representation:
$$ e^{-x_3^2} H_{n}(x_3) \sim \vert n \rangle =
{{(a^{\dagger})^{n}}\over{ \sqrt{n!}}} \vert 0 \rangle, \quad a =
\left( {\half} {\p}_3 + x_3 \right), a^{\dagger} = \left( -
{\half} {\p}_3 + x_3\right) . $$ Then:
\eqn\ovrlp{\eqalign{{\CV}_{\Vert} (k, n,
{\nb})& = \langle {\nb} \vert e^{-{ik \over{2}} ( a +
a^{\dagger})}\vert n \rangle \cr & = e^{-{k^2
\over{8}}} \sum_{m \geq 0} \left( {{i k}\over{2}}\right)^{{\nb} +
n - 2m} {{\sqrt{n! {\nb}!}}\over{m! ( n - m)! ({\nb} -m)!}}\ , \cr
{\CV}_{\perp} ({\kk}, n, {\nb}) &= e^{-{{\kk{\bar\kk}}\over{2}}}
\sum_{\scriptstyle{m \geq 0}} {{ \sqrt{n! {\nb}!} \,
(-i{\kk})^{{\nb}-m} (-{\bar\kk})^{n-m}}\over{m! ( n -m)!
({\nb}-m)!}} \ .\cr}}

The three-point disc amplitudes are labelled by the types of the
D-brane boundary conditions one imposes between the points of
operator insertions. Up to cyclic permutations we have the
following options: $$1-1-1, 1-1-3, 1-3-3, 3-3-3\ . $$ They correspond
to the triple vertices between the following modes in the gauge
theory: $$ a^{11}a^{11}a^{11}, a^{11}a^{13}a^{31},
a^{13}a^{33}a^{31}, a^{33}a^{33}a^{33}\ ,$$ respectively. We shall
look at the bosonic vertex only, the ${\l}{\l}a$ vertices follow
 by supersymmetry.

The term ${\CL}_3$ being proportional to commutators will vanish
for $N=1$ in the $1-1-1$ sector.

For other sectors it is convenient to rewrite the ${\Tr} D_{B}^0
a_{C} [ a_{B}, a_{C}] $ term
 as $$ 2 {\Tr} (D_{B}^{0} a_{C})
a_{B}a_{C}\ ,$$ using the gauge condition \gcn.

After straightforward computation we arrive at the following
amplitudes $$ {\CA} = {\d} ({\o}_1 + {\o}_2 - {\o}_3)
{\CA}^{abc}({\o}_1, {\o}_2, {\o}_3)$$
\eqn\trampl{\eqalign{ &
{\CA}^{113} = 2i{\CV}_{\Vert} (k_1, n, l) \quad \times [ \cr &
{\o}_1 \, {\z}_A ({\o}_1, k_1) \left(  {\bar\z}_A ({\o}_3 , m, l)
{\z}_t ({\o}_2, m,n) - {\z}_{A} ({\o}_2 , m,n) {\bar\z}_t ({\o}_3,
m,l) \right) -\cr &  k_1 \, {\z}_A ({\o}_1, k_1) \left( {\bar\z}_A
({\o}_3 , m, l) {\z}_3 ({\o}_2, m,n) - {\z}_{A} ({\o}_2 , m,n)
{\bar\z}_3 ({\o}_3, m,l) \right) - \cr & {\o}_2 {\z}_t ({\o}_1,
k_1)  {\z}_A ({\o}_2, m,n) {\bar\z}_A ({\o}_3, m,l) + \cr i
\sqrt{m} & \left( {\z}_1 + i {\z}_2 \right) ({\o}_1, k_1) {\z}_A
({\o}_2, m+1, n) {\bar\z}_{A} ({\o}_3, m, l) - \cr i \sqrt{m} &
\left( {\z}_1 - i {\z}_2 \right) ({\o}_1, k_1 ) {\z}_A ({\o}_2,
m,n) {\bar\z}_{A} ({\o}_3, m+1, l) -\cr i \sqrt{n+1} & \left(
{\z}_3 + i {\z}_4 \right) ({\o}_1, k_1) {\z}_A ({\o}_2, m, n+1)
{\bar\z}_{A} ({\o}_3, m,l) +\cr i \sqrt{n} & \left( {\z}_3 + i
{\z}_4 \right) ({\o}_1, k_1) {\z}_A ({\o}_2, m, n-1) {\bar\z}_A
({\o}_3, m,l) ] \cr & {\CA}^{133} = 2i {\CV}_{\Vert}(k_1, n,l)
{\CV}_{\perp} ({\kk}_1, m-1, {\mb} -1) \quad \times  [\cr &
{\z}_{A}({\o}_1, {\vec k}_1)  k_{1}^{\m} \left( {\bar\z}_{\m}
({\o}_2, m, l) {\z}_{A} ({\o}_3, {\mb}, n)  - {\z}_{\m}({\o}_3,
{\mb}, p) {\bar\z}_{A} ({\o}_2, m, l) \right) \cr &  {\o}_2
{\z}_{t} ({\o}_1, {\vec k}_1) {\z}_A ({\o}_2, {\mb}, n)
{\bar\z}_{A} ({\o}_3, m, l) + \cr i\sqrt{{\mb}}& \left( {\z}_1 + i
{\z}_2 \right) ({\o}_1, {\vec k}_1) {\z}_A ({\o}_2, {\mb}+1, n)
{\bar\z}_A ({\o}_3, m, l) -\cr i\sqrt{{\mb}-1}& \left( {\z}_1 - i
{\z}_2 \right) ({\o}_1, {\vec k}_1) {\z}_A ({\o}_2, {\mb}-1, n)
{\bar\z}_A ({\o}_3, m, l) -\cr
 i \sqrt{n} & \left( {\z}_3 + i {\z}_4 \right)
({\o}_1, {\vec k}_1) {\z}_{A} ({\o}_2, {\mb}, n-1) {\bar\z}_{A}
({\o}_3, m, l)  - \cr i \sqrt{n+1} & \left( {\z}_3 - i{\z}_4
\right) ({\o}_1, {\vec k}_1) {\z}_{A} ({\o}_2, {\mb}, n+1)
{\bar\z}_{A} ({\o}_3, m, l)] \ . \cr } }
Finally, the bulk +
bulk$\to$ bulk amplitude coincides with that of the ordinary
$U(1)$ \nc gauge theory:
\eqn\trtrtr{\eqalign{{\CA}^{333} = -
{\d}^{(3)}({\vec k}_1 + {\vec k}_2 + {\vec k}_3) \sin
\left(-{1\over 2} {\vec k}_2 \times {\vec k}_3 \right) k_{1}^{\m}
{\z}_A ({\o}_1, {\vec k}_1) {\z}_{\m} ({\o}_2, {\vec k}_2)
{\z}_{A} (- {\o}_3, {\vec k}_3) \ . \cr}}

 \subsec{The quartic
interactions} The quartic vertices arise from the term
\eqn\qutr{S_{4} = - {{\pi g^2} \over 2} \int {\rm d} t {\rm d} x_3
{\Tr} \left( a_{A} a_{B} a_{A} a_{B} - a_{A}^2 a_{B}^2 \right)\ .}
We  substitute \dcmps\ into \qutr\ to get the vertices for the
quartic interactions between the various quanta ${\z}_{A}$.
Basically all we need to do is to multiply \prdc\ by \prdc,
shuffle the indices around and to take the trace over $\CH$. The
resulting expression is rather messy. We present here only the
amplitudes which are responsible for $$ 11 + 11 \to 13 + 31 $$
process and those related to it by  crossing: ${\d} ( {\o} +
{\o}^{\prime} + {\tilde\o} + {\tilde\o}^{\prime} ) {\CA}_4 ({\o},
\ldots) $, where:
\eqn\qrtcvrt{\eqalign{ & {\CA}_4 = \left[
{\z}_{A} ({\tilde\o}, {\tilde n}, m) {\bar\z}_{B} ( -
{\tilde\o}^{\prime}, {\tilde n}^{\prime}, m ) - {\bar\z}_{A} (-
{\tilde\o}, {\tilde n}, m) {\z}_{B} ( {\tilde\o}^{\prime}, {\tilde
n}^{\prime}, m ) \right] \times \cr &  {\z}_{A} ({\o}, k) {\z}_{B}
({\o}^{\prime}, k^{\prime}) {\CV}_{113} ( k + k^{\prime}, {\tilde
n}, {\tilde n}^{\prime}) + \cr & (\left[ {\z}_{A} ({\o}, k)
{\z}_{B} ({\o}^{\prime}, n, m) - A \leftrightarrow B \right]
{\z}_{A} ({\tilde\o}, {\tilde k}) {\bar\z}_{B}
(-{\tilde\o}^{\prime}, {\tilde n}, m)  + \cr & \left[ {\z}_{A}
({\o}, k) {\bar\z}_{B} (-{\o}^{\prime}, n, m) - A \leftrightarrow
B \right] {\z}_{A} ({\tilde\o}, {\tilde k}) {\z}_{B}
({\tilde\o}^{\prime}, {\tilde n}, m)) {\CV}_{113}( k + {\tilde k},
n, {\tilde n})\ . \cr}}

\newsec{Fluctuations and interactions of the $N$-fluxon}

In this section we shall briefly discuss the generalization
of the above discussion of the fluctuations and interactions of the fluxon
to the case of the $N$-fluxon. In this case the Hilbert space is decomposed into
\eqn\spltt{ {\CH} = V_{N} \oplus {\CH}_{N},  \quad V_{N}=P_N\CH, \, {\CH}_{N}
=(1-P_N){\CH}  \ , }
and the background fields satisfy
$A_4^0=i\Phi^0, \, F_{1,2}^0, \, F_{13,4}^0 \propto P_N$
and $D, \, \bar D \in {\CH}_{N}$.
Thus, as before, we shall write the  space ${\rm End}({\CH})$ of the
operators in the Fock space    as a direct sum of subspaces:
\eqn\spltt{{\rm End}{\CH} = V_{11} \oplus V_{13} \oplus V_{31}
\oplus V_{33}\ , }
where
\eqn\sbspt{V_{11} = V_{N} \otimes V_{N},
\quad V_{13} = V_{N} {\otimes} {\CH}_{N}, \quad V_{31} = {\CH}_{N}
\otimes V_{N}, \quad V_{33} = {\CH}_{N} \hat\otimes {\CH}_{N}\ .}

Let us consider the 1-1 modes that are described by $A_B=A_B^0+g
a_B$, where $a_B \in V_{11}$, and write these as \eqn\nflcomp{a_B
= {1\over\sqrt{\t}} \sum_{i,j=1}^N {\CA}_{B}^{ij}\vert
i\rangle\langle j \vert, \quad B=0,1, \dots 9\, , \quad {\bf
\CA}_{B}= [\CA_{B}^{ij}] .} Notice that we have normalized ${\bf
\CA}_{B}$   to be dimensionless, as befits a two-dimensional gauge
field. Then, using the fact that $[D,a_B]=[\bar
D,a_B]=[F_{AB},a_C]=0$, we easily find that:
\eqn\lagN{\eqalign{\CS=\int
dtdx_3 &\left[ {2\pi\over g^2 \t}  + {\rm Tr}\left(  {1\over
2}\CA_B(\p_t^2-\p_3^2)\CA_B +{g \over
\sqrt{\t}}\p_\a\CA_B[\CA_\a,\CA_B]+  {g^2 \over \t}[\CA_B,\CA_C]^2
\right)\right] \cr + & {\rm fermions}  \ , \cr}}
where $\a = 0,3$ and we are working in the gauge $\p_t\CA_0-\p_3\CA_3=0 .$
Thus, as expected, the 1-1 fluctuations about the $N$-fluxon
are described by the 2 dimensional $U(N)$ ${\CN}=8$
super-Yang-Mills theory, the dimensionally reduced d=10
supersymmetric gauge theory, with coupling ${g / \sqrt{\t}}.$

If we were to perturb about the separated $N$-fluxon \higss\ , the above
action for the 1-1 modes would be modifed, since now $\Phi =-2x_3P_N +D_N$ is
no longer proportional to the identity in $V_N$. In particular,  we will generate
from the quartic term in the action a mass term for the 1-1 modes equal to
$\sum_{ij}(d_i-d_j)^2\CA_{ij}^2$ , as befits the low energy modes of  fundamental
strings
on separated D1 branes.

In addition, one can easily derive, following the steps described
above for the 1-fluxon, the spectrum and interactions of the
1-3,3-1, and 3-3 modes. Note that the 1-3 and 3-1, discrete
energy,  modes will now be in the fundamental representation of
$U(N)$.

\newsec{Dyons, ${\vartheta}$-angles}

So far we have been studying   magnetic flux tubes, which were
obtained by solving the Bogomolny equation $$[D_i, {\Phi}] + B_i =
0\ . $$ Let us look at more general solutions, with both electric
and magnetic charges. The appropriate BPS equations are gotten by
using the \nc analogue of the usual BPS inequalities:
\eqn\ineq{\eqalign{g^2{\vec E}^2 + {1\over{g^2}} {\vec B}^2
+{1\over{g^2}} ({\vec D}\Phi)^2 & = (g {\vec E}+{{\sin\a}\over{g}}
\, {\vec D}\Phi)^2 +{1\over{g^2}} ({\vec B}+\cos\a \, {\vec
D}\Phi)^2- \cr & ({\vec E} \sin\a + {1\over{g^2}} {\vec B} \cos\a)
\cdot {\vec D}\Phi - {\vec D}\Phi \cdot ( {\vec E} \sin\a +
{1\over{g^2}} {\vec B} \cos\a) \geq \cr -  {\vec D} \cdot & \left(
{\Phi} \star \left( \sin\a \, {\vec E} + {{\cos\a}\over{g^2}} \,
{\vec B} \right)  + \left( \sin\a \, {\vec E} +
{{\cos\a}\over{g^2}} \, {\vec B} \right) \star \Phi \right) \ .
\cr }} They are:
\eqn\embps{\eqalign{ {\rm cos}\, {\a} \,\, [D_i,
{\Phi}] + B_i &= 0\ ,\cr
 {\rm sin}\, {\a} \,\, [D_i, {\Phi}] + g^2
E_i &= 0\ , \cr}}
where $E_i = -i {{F_{0i}}\over{g^2}}$ is the
electric field. They are easy to solve: take $A_i$ to be equal to
the monopole solution gauge field $A_i^{\rm mon}$ found in
\grossnek, or to the $q$-fluxon solution $A^{\rm flux}$ discussed
in this paper, and then take
\eqn\scan{\eqalign{\Phi =
{{\Phi}^{\rm mon} \over {\cos\a}}\quad & {\rm or} \quad
{{\Phi}^{\rm flux} \over {\cos\a}}\ ,\cr A_t = - i \, {\tan\a} \,\,
{\Phi}^{\rm mon} \quad & {\rm or} \quad - i \, {\tan\a} \,\,
{\Phi}^{\rm flux}, \quad A_3=0 \ .\cr }}

In this way one gets a solution with both electric and magnetic
charges: \eqn\chrgs{g^2 \, Q_{e} = {\tan\a}\, Q_{m}=q\, {\tan\a}\
. } The magnetic field will be as before (dyons in the noncommutative 
Yang-Mills theory were also recently 
discussed in \baklee, note the $2\pi$ factor
difference in our normalization of electric charge compared to that 
reference). 

Consider the case of the $q$-fluxon. The only nonvanishing
component of the electric field will be \eqn\elec{g^2 E_3= \,
{\tan\a} \,  B_3 \ ,} so that we have a flux tube of magnetic and
electric field along the $x_3$ axis. The tension of this dyonic
fluxon will be
\eqn\tendy{T={2q\pi\over g^2 \t(\cos\a)^2}\ .}
Of course, in the quantum theory $Q_{e}$ must be an integer. This
will emerge in the standard fashion once we quantize the solitons.
This will fix the allowed values of ${\a}$ so that $$\tan \a
=g^2{p\over q}, \quad p= 1,2,3,...$$ and the tension will be
\eqn\tend{T={2 \pi\over q\t}\left({{q^2}\over{g^2}}+ g^2 p^2
\right)\ .} This can be seen by transforming the above solution to
$A_t=0$ gauge, wherein $D, \bar D, $ are unchanged and
\eqn\Atgauge{A_t=0, \quad \Phi =-2x_3P_N, \quad A_3 =  -2it\times
{\tan\a} P_N \quad .} Let us briefly discuss how this formula is
consistent with the familiar $S$-duality covariant expression for
the tension of the $(p,q)$ string:
\eqn\tpq{T_{(p,q)} =
{1\over{\ap}} \sqrt{{{q^2}\over{(2\pi g_{s})^2}} + p^2}\ .}
In the presence of the $\vartheta$-angle the relation between the
electric and magnetic charges is modified. The $\vartheta$-angle
is the vev of the ten-dimensional axion field, which makes the
dilaton a complex field.  It will effectively shift $p$ by
${\vartheta}q$ in the formula above. The same effect is present in
the gauge theory (Witten's effect). Let us therefore set
${\vartheta}= 0$ for clarity.

Recall \witsei\ the relation between the closed string background
and the parameters of the \nc gauge theory. We have the $B$-field
${\half}B {\rm d}x^1 \wedge {\rm d} x^2$, closed string coupling
$g_{s}$ which entered \tpq:
\eqn\dict{g_{s} = {{g^2
{\ap}}\over{\sqrt{(2\pi \ap)^2 + {\t}^2}}}, \quad B =
{{\t}\over{(2\pi \ap)^2 + {\t}^2}}\ .} Recall that the $B$-field
tilts the D-string towards the D3 brane
\moriyama\hashimoto\grossnek. From our formulae for the scalar
field \scan\ and  the Dirac-Born-Infeld intuition as in \grossnek\
we conclude that the $(p,q)$-string forms an angle
${\psi}_{(p,q)}$ with the D3 brane:
\eqn\pqan{{\tan}{\psi}_{(p,q)}
= {{2\pi\ap}\over{{\t} \, {\cos\a}}}\ .} The tension of the
projection onto the D3 brane  of the tilted $(p,q)$-string is
equal to:
\eqn\tnprj{{{T_{(p,q)}}\over{\cos\psi_{(p,q)}}} =
{{2\pi}\over{g^2 {\t}q}} \left( q^2 + g^4 p^2 \right) + {\Delta}T\ , }
with $$ {\Delta}T = {{q {\t}}\over{2\pi (g \ap)^2}}$$ being the
defect. Notice that this defect term depends only on the magnetic
charge. It is $q$ times higher then the analogous defect term
computed in \grossnek\ where it was interpreted as the work done
by the magnetic force in bringing the semi-infinite string in from
infinity. We should imagine constructing our dyonic string by
taking the semi-infinite string, the $(p,q)$-version of the
solution of \grossnek, and then translating it as in the section
3.2.  Although the endpoint of the $(p,q)$ string is a $(p,q)$
dyon, the background space-like $B$ field couples only to the
magnetic charge, i.e. only to $q$. Upon subtracting this defect we
 precisely match the gauge theory answer \tend\
and the tension expected from  the D-brane consideration.

\newsec{Conclusions}

In this paper we have constructed a large class of very simple
BPS solutions of d=4, supersymmetric \nc Yang-Mills theory---that correspond
to D1 strings intersecting D3 branes in the presence of a background B-field in the
decoupling limit. We analysed in some detail the fluctuations about
these solitons and their interactions. The results were in complete
agreement with the expecteations from string theory. In particular
we found the fluctuations of the superstring in 10
dimensions arising from fundamental strings
attached to the D1 strings, the ordinary particles of the gauge theory in 4
dimensions and a set of states with  discrete spectrum, localized at the
intersection point,  corresponding to fundamental strings stretched between the  D1
string and the D3 brane.

    Unfortunately, in the semiclassical treatment that we have employed we are
unable to
see the massive modes of the D1 string, whose energies are of order $1/g^2$. Perhaps
some of these modes are visable as BPS states that can be constructed as
gauge theory solitons.

There are many facets of our investigation which remain to be completed;
including the explicit construction of \nc monopole strings for
\nc $U(N)$ theory and the generalization of our considerations to other branes and
dimensions.
For example, it is easy to use of construction of fluxons to construct vortex
solitons of 1+2 dimensional \nc gauge theory. Simply take our solution
for $\bar D$  and $D$, \nflx, and throw away the $\Phi$ field.

Finally, the implications of these solitons for the dynamics of large N-gauge theory
remain to be investigated.

\bigskip
\bigskip
\bigskip
\ndt {\bf Acknowledgements.}

\ndt{}We would like to thank  A. Hashimoto, N. Itzhaki,
G. Moore, Polchinski and  E. Witten for
discussions. Our research was partially supported by NSF under the
grants PHY94-07194 and PHY 97-22022; in addition, research of NN was supported by
Robert H.~Dicke fellowship from Princeton University, partly by
RFFI under grant 00-02-16530, partly by the grant 00-15-96557 for
scientific schools. NN is grateful to ITP, UC Santa Barbara and IHES
at  Bures-sur-Yvette, for their hospitality during various stages
of this work.

\footatend\vfill\supereject\immediate\closeout\rfile\writestoppt
\baselineskip=14pt\centerline{{\bf References}}\bigskip{\frenchspacing%
\parindent=20pt\escapechar=` \input refs.tmp\vfill\eject}\nonfrenchspacing \bye